\documentclass[12pt]{article}
\usepackage{amsmath}
\usepackage{amsmath,bm}
\usepackage{url}
\usepackage{tikz}
\usetikzlibrary{shapes.geometric}
\usetikzlibrary{arrows}
\usepackage{amsmath}
\usepackage{amssymb}
\usepackage{hyperref}
\usepackage{rotating}
\usepackage{physics}
\usepackage{epsfig}
\usepackage{graphicx}
\usepackage{color}
\usepackage{cite}
\usepackage{subfig}
\usepackage{hhline}
\usepackage[font=small,labelfont=bf]{caption}

\newcommand{\beq}{\begin{equation}}
\newcommand{\eeq}{\end{equation}}
\newcommand{\ga}{\lower.7ex\hbox{$\;\stackrel{\textstyle>}{\sim}\;$}}
\newcommand{\la}{\lower.7ex\hbox{$\;\stackrel{\textstyle<}{\sim}\;$}}

\hypersetup{
    colorlinks = true,
    citecolor = {blue},
    linkcolor = {blue},
    urlcolor = {blue},
}

\numberwithin{equation}{section}

\begin{document}

\def\jcap{\ref@jnl{J. Cosmology Astropart. Phys.}}

\begin{flushright}
{\tt UMN-TH-4207/23, FTPI-MINN-23/01} \\
{\tt CPHT-RR001.012023} \\
\end{flushright}

\vspace{0.2cm}
\begin{center}
{\bf {\Large Supergravity Scattering Amplitudes }}

\end{center}
\vspace{0.1cm}

\begin{center}{
{\bf Emilian Dudas}$^{a}$,
{\bf Tony Gherghetta}$^{b}$,
{\bf Keith~A.~Olive}$^{b,c}$, and
{\bf Sarunas~Verner}$^{d}$}
\end{center}

\begin{center}
{\small\em $^a$CPHT, CNRS, {\'E}cole Polytechnique, Institut Polytechnique de Paris, 91120 Palaiseau, France}\\
{\small\em $^b$School of
 Physics and Astronomy, University of Minnesota, Minneapolis, MN 55455,
 USA} \\
{\small\em $^c$William I. Fine Theoretical Physics Institute, School of
 Physics and Astronomy, University of Minnesota, Minneapolis, MN 55455,
 USA} \\
 {\small\em $^d$Institute for Fundamental Theory, Physics Department, University of Florida, Gainesville, FL 32611, USA}
 \end{center}

\vspace{0.1cm}
\centerline{\bf Abstract}
\vspace{0.1cm}

{\small Supergravity theories with non-minimal K\"ahler potentials are characterized by a non-trivial field space manifold with corresponding non-trivial kinetic terms. 
The scattering amplitudes in these theories can be calculated at fixed background field values 
by making a field redefinition to Riemann normal coordinates. Because of the K\"ahler structure of supergravity, a more compact method for calculating amplitudes is obtained by a redefinition to K\"ahler normal coordinates. 
We compare both methods and calculate the explicit transformations and amplitudes for several examples in the context of no-scale supergravity with one and two chiral superfields. We show that in all cases the equivalence of the scattering amplitudes using either Riemann normal or K\"ahler normal coordinates is possible only at extremal points of the scalar potential.} 

\vspace{0.2in}

\begin{flushleft}
{February} 2023
\end{flushleft}
\medskip
\noindent

\newpage

\section{Introduction}
A reasonable selection criterion in the search for viable theories encompassing physics beyond the Standard Model, is a UV completion which is derivable from a quantum theory of gravity, e.g., string theory.
The calculation of scattering amplitudes has attracted considerable attention because of the many conjectures which demand scattering amplitudes larger than gravitational amplitudes which is postulated to be the weakest interaction.  For example, the Weak Gravity Conjecture \cite{ArkaniHamed:2006dz}
places constraints on the gauge coupling of a U(1) gauge symmetry to ensure the stability of extremal black holes. A violation of the conjecture could be interpreted as the incompatibility of the gauge theory with a proper UV completion stemming from string theory. As such, the theory would be designated as part of the swampland \cite{Vafa:2005ui}.

A more ambitious conjecture known as the Scalar Weak Gravity Conjecture (SWGC) \cite{Palti:2017elp} further speculates that for all scalar fields mediating interactions, there must exist a state for which the scalar exchange is stronger than gravity. A stronger version known as the Strong Scalar Weak Gravity Conjecture \cite{Gonzalo:2019gjp} states that the latter must be true for all values of the scalar fields and can be encapsulated by the limit \cite{Gonzalo:2019gjp,Benakli:2020pkm}
\beq
m^2 \frac{\partial^2}{\partial \phi^2}\left(\frac{1}{m^2}\right) \ge \frac{1}{M_P^2} \, ,
\eeq
where $m^2 = \partial^2 V/\partial \phi^2$ is the effective mass of the scalar field, $\phi$, and $M_P \simeq 2.4 \times 10^{18} \, \rm{GeV}$ is the reduced Planck mass. There have since been many applications of these conjectures on various scalar field theories, including inflation \cite{Scalisi:2018eaz,Gonzalo:2019gjp,Kusenko:2019kcu,Shirai:2019tgr,DallAgata:2020ino,Andriot:2020lea,Benakli:2020pkm}.

Many of the examples considered are theories with 
a single scalar field, $\phi$, and a canonical kinetic term. In this case, testing the conjecture amounts to a calculation of the four-point scattering amplitude $\phi \phi \to \phi \phi$, which may receive contributions from a contact (quartic) term in the potential, as well as $s$, $t$, and $u$ channels of $\phi$-exchange arising from cubic interactions. The resulting amplitude can then be directly compared with the gravitational scattering amplitude. 

In theories with multiple scalar fields, complex geometries may arise where the scalar fields act as coordinates on a background field-space manifold. Focusing on specific directions in field space may not capture the full set of scalar field interactions, even when the fields are canonically redefined. In general, canonical field redefinitions are only possible for a given  fixed background. As in general relativity, coordinate transformations to flat Minkowski space-time amount to transformations to Riemann normal coordinates valid at a specific space-time point. Analogously, the calculation of scattering amplitudes in non-flat field space geometries are most directly performed in terms of Riemann normal field definitions at a specific background field value.

Similarly, theories of supergravity are characterized by a non-trivial metric for complex manifolds that arises from the scalar fields of chiral multiplets. Typically, such theories have the usual quadratic kinetic terms in the action, and are described by a complex K\"ahler manifold. In general, one can perform a holomorphic transformation to K\"ahler normal coordinates~\cite{Higashijima:2000wz, Higashijima:2002fq} that ensures that the K\"ahler metric is flat at a specific space-time point.\footnote{We note that K\"ahler normal coordinates are also known as canonical coordinates.} In these coordinates the kinetic terms of the Lagrangian are canonical and there are no cubic order terms. The absence of cubic terms make these coordinates particularly useful when calculating the scattering amplitudes.

A common framework for non-trivial field space geometries is
${\cal N} = 1$ supergravity theories \cite{Nilles:1983ge}, where the metric for the complex field space manifold is given 
by 
\beq
\label{metric1}
g_{i \bar{j}} \; = \; K_{i {\bar j}} \; \equiv \; \frac{\partial^2 K}{\partial Z^i \partial {\bar Z^{\,\bar{j}}}} \; \equiv \; \partial_{i} \partial_{\,\bar{j}} K \, ,
\eeq
where $Z$ is a complex scalar field, $K(Z^i, {\bar Z^{\,\bar{i}}})$ is the K\"ahler potential, and the inverse of the K\"ahler metric is given by $K^{i \bar{j}}$. We follow the convention where the unbarred (holomorphic) indices are on the left and barred (anti-holomorphic) indices are on the right. Therefore, $K^{i \bar{j}} K_{m \bar{j}} = \delta^i_m$. We also note that the metric~(\ref{metric1}) and all of the $\mathcal{N} = 1$ supergravity action is invariant under the K\"ahler transformation
\begin{equation}
    \label{kahtrans}
    K(Z^i, {\bar Z^{\,\bar{i}}}) \longrightarrow K(Z^i, {\bar Z^{\,\bar{i}}}) + f(Z^i) + \bar f(\bar{Z}^{\,\bar{i}}) \quad , \quad W \longrightarrow e^{-f(Z^i)}  W  \ , 
\end{equation}
where $f(Z^i)$ is an arbitrary holomorphic function and $W(Z^i)$ is the superpotential and is also holomorphic.

The natural framework for a low-energy field theory derived from string theory \cite{Witten} is no-scale supergravity \cite{no-scale}.
The simplest $\mathcal{N} = 1$ no-scale supergravity models were first considered in~\cite{no-scale, Ellis:1983sf}, and are characterized by the following K\"ahler potential~\cite{EKN1}:
\begin{equation} 
\label{kah1}
K \; = \; - \, 3 \alpha \ln (T + \bar{T}) \, ,
\end{equation}
where $T$ is a complex chiral field that can be identified as the volume modulus fiel,\footnote{ We work in units of the reduced Planck mass, $M_P = 1/\sqrt{8 \pi G_N}$. In some cases, to avoid any potential confusion, explicit factors of $M_P$ will be included.} $\bar{T}$ is the conjugate field, and $\alpha$ is a real parameter associated with the curvature of the K\"ahler manifold.  The minimal no-scale K\"ahler potential (\ref{kah1}) describes a non-compact $\frac{SU(1,1)}{U(1)}$ coset manifold. This can be generalized by including matter fields $\phi^i$ that parametrize, together with the volume modulus $T$, an $\frac{SU(N,1)}{SU(N)\times U(1)}$ coset space, 
defined by the K\"ahler potential~\cite{EKN}
\beq
\label{v0}
K \; = \; -3  \alpha \ln (T + \bar{T} - \frac{|\phi^i|^2}{3}) \, .
\eeq
There are other generalizations based on other non-compact coset spaces, which also appear in some string models and involve multiple moduli fields. 
For example, testing the SWGC for inflationary models derived from no-scale supergravity \cite{building} would require inclusion of all scalar interactions between the inflaton and moduli fields.

In this paper, we apply the formalism for calculating scattering amplitudes in no-scale supergravity models. However, we note that the formalism presented here is completely general and can be applied to general scalar or supergravity theories. We begin in Section \ref{general} by outlining the procedure for the necessary field redefinitions to Riemann normal scalar fields in a given background. In this basis, we provide general expressions for the three- and four-point scattering amplitudes. We also introduce the field redefinitions to K\"ahler normal coordinates in supersymmetric theories. Scattering amplitudes in terms of K\"ahler normal coordinates are also provided. These are applied to supergravity theories in Section~\ref{sugr}. 

In Section \ref{su11}, we consider the simplest no-scale models based on the $\frac{SU(1,1)}{U(1)}$ coset manifold. In the absence of a superpotential for $T$, we quickly see that two of the four-point amplitudes vanish (directly contradicting the SWGC). We then consider a toy extension, where an ``effective mass" term of the form $m^2 T {\bar T}$ is included. Despite the apparent form of the potential, this theory has no stable vacua, and as a consequence, the amplitudes in Riemann and K\"ahler normal coordinates do not agree at any point in field space. 
This is next followed with two more interesting examples. One is derived from a simple superpotential, $W = m(T-1)^2$ which gives two extrema (a minimum and a maximum). We compute again the scattering amplitudes in both field bases, and show explicitly the invariance of the amplitude under these field redefinitions 
solely at the two extrema. The second example is
based on the KKLT model \cite{Kachru:2003aw} for moduli stabilization (with and without an uplift). Without an uplift, the potential has only one extremum, a minimum, and with the uplift, there are two extrema. Once again we show that the field redefinition invariance of the amplitudes is achieved only at these extrema.  The invariance holds also for any value of  fields along a flat direction, since there is no a linear term (tadpole) along the flat direction. Then in Section \ref{su21}, we provide the formalism for a two-field model based on Eq.~(\ref{v0}). Here too, we consider examples with and without a superpotential.  We summarize our work in Section \ref{summary}.

\section{Scattering Amplitudes with General Two-Derivative Interactions and Potential}
\label{general}
We begin our discussion by briefly summarizing the geometric structure and properties of scalar field theories. In particular, we show how to use the geometry of the field space to calculate the three- and four-point scattering amplitudes. We extend the well-known expressions for the three- and four-point scattering amplitudes in terms of geometric invariants~\cite{Alonso:2015fsp,Cohen:2021ucp,Alonso:2021rac,Cheung:2021yog,Cheung:2022vnd,Helset:2022tlf} and include the relevant mass terms. We first discuss theories of $N$ massive scalars with two-derivative interactions using the Riemann normal coordinates and then introduce the analog of K\"ahler normal coordinates in supersymmetric theories. Importantly, we show that the scattering amplitudes are independent on the choice of coordinates provided that they are computed at an extremum of the potential, i.e.  with an absence of tadpoles.

\subsection{Coordinate Transformations}
\subsubsection{Real scalar fields}
Consider a theory of $N$ massive real scalars $\Phi^I$ with general two-derivative interactions and an arbitrary potential $V(\Phi)$. The Lagrangian is given by
\begin{equation}
    \label{lag1}
    {\cal L} \; = \;\frac{1}{2} g_{IJ}(\Phi)\partial_\mu \Phi^I \partial^\mu \Phi^J - V(\Phi) \, ,
\end{equation}
where $g_{IJ}(\Phi)$ is an arbitrary symmetric metric.  We use the uppercase Latin indices $I, J, K, \ldots,$ to denote the \textit{flavor eigenbasis}.\footnote{We refer to the flavor eigenbasis even in the absence of the scalar potential. However, since the choice of the metric $g_{IJ}$ is completely arbitrary, we also refer to the states that are not canonically normalized as flavor eigenstates.} 

In general, the coordinate transformation (field redefinition) $\Phi^I \rightarrow \Phi'^I$ implies the following scalar, vector, and tensor transformation laws:
\begin{eqnarray}
    \label{transflaws}
    V(\Phi) &\rightarrow&  V(\Phi') \, , \\
    \partial_{\mu} \Phi^I &\rightarrow&  \frac{\partial \Phi'^I}{\partial \Phi^J} \partial_{\mu} \Phi^J \, ,\\ 
    g_{IJ} (\Phi) &\rightarrow& \frac{\partial \Phi^K}{\partial \Phi'^I} \frac{\partial \Phi^L}{\partial \Phi'^J} g_{KL}(\Phi') \, .
\end{eqnarray}
Note that the on-shell scattering amplitudes arising from the Lagrangian~(\ref{lag1}) must be invariant under the coordinate transformation $\Phi^I \rightarrow \Phi'^I$ in a well defined theory. Therefore, one can always find a set of unique coordinate transformations that make an arbitrary metric $g_{IJ}(\Phi)$ flat at a particular point in field space.

To calculate the physical scattering amplitudes, we will, for now, consider only background field values corresponding to stable vacua. We introduce the following field expansion
\begin{equation}
    \Phi^I \; = \; v^I + \phi^I \, ,
\end{equation}
where $v^I$ is the vacuum expectation value (VEV) of the scalar field and $\phi^I$ is the fluctuation of the scalar field about the stable vacuum. Therefore, all physical interactions are governed by the Lagrangian couplings and the corresponding derivative couplings of the dynamical field fluctuations $\phi^I$ evaluated at the VEV. As we shall see, on-shell scattering amplitudes are invariant under coordinate transformations when the chosen vacuum is stable, which implies that the tadpole $\partial_I V(v)$ vanishes. We show in Section~\ref{su11} that the on-shell scattering amplitudes are in fact no longer invariant when evaluated at a non-extremal point in field space.

Following Ref.~\cite{Cheung:2021yog}, we introduce the symmetrized covariant derivatives of the  potential:
\begin{equation}
    V_{I_1 \dots I_k} (v) \; = \; \nabla_{(I_1\dots} \nabla_{I_k)} V(\Phi)|_v\,,  \quad {\rm{with}} \quad \partial_I V(\Phi)|_v \; = \; 0 \, ,
\end{equation}
where the symmetrized covariant derivative includes the sum of all permutations multiplied by a symmetry factor of $1/n!$. We note that the derivatives are taken with respect to the unshifted fields and evaluated at their VEVs, $v^I$, and not the dynamical field fluctuations $\phi^I$. The covariant derivative of a general tensor is defined as
\begin{equation}
    \begin{aligned}
    \label{eq:gentender}
\nabla_{M} T^{I_1 I_2 \cdots I_k}{ }_{L_1 L_2 \cdots L_n}  & =  \partial_M T^{I_1 I_2 \cdots I_k}{ }_{L_1 L_2 \cdots L_n}  + \Gamma_{MN}^{I_1} T^{N I_2 \cdots I_k}{ }_{L_1 L_2 \cdots L_n}+\Gamma_{MN}^{I_2} T^{I_1 N\cdots I_k}{ }_{L_1 L_2 \cdots L_n}+\cdots\\
&-\Gamma_{M L_1}^N T^{I_1 I_2 \cdots I_{k}} { }_{N L_2 \cdots L_n}-\Gamma_{M L_2}^N T^{I_1 I_2 \cdots I_{k}} { }_{L_1 N \cdots L_n}-\cdots \, ,
\end{aligned}
\end{equation}
where the Christoffel symbols are given by
\begin{equation}
    \label{christofsymbols}
    \Gamma^I_{JK} \; = \; \frac{1}{2} g^{IL} \left(g_{KL,\,J} + g_{LJ,\,K} - g_{JK,\,L} \right) \, .
\end{equation}

From these definitions, the mass matrix evaluated at an extremal vacuum point is given by $V_{IJ}(v) = \partial_I \partial_J V(\Phi)|_v$. Therefore, from the Lagrangian~(\ref{lag1}) we find that the linearized equation of motion in the \textit{flavor eigenbasis} is given by
\begin{equation}
    \left(g_{IJ}(v)\Box + V_{IJ}(v) \right) \phi^I(x) \; = \; 0 \, ,
\end{equation}
where $\Box \equiv \partial^{\mu} \partial_{\mu}$ is the d'Alembertian. 
Our next goal is to canonically normalize and diagonalize the linearized equation of motion and transform from the \textit{flavor eigenbasis} to the \textit{mass eigenbasis}. This can be readily done by flattening an arbitrary metric $g_{IJ}$ using the coordinate transformations~(\ref{transflaws}):
\begin{equation}
    g_{IJ}(v) \frac{\partial \Phi^I}{\partial \Phi'^i} \Big|_v \, \frac{\partial \Phi^J}{\partial \Phi'^j} \Big|_v \; = \; \delta_{ij} \, .
\end{equation}
Equivalently, we can introduce the vielbein field that flattens the metric:\footnote{In the physics literature, vielbein is also referred to as a vierbein or tetrad. However, since we consider an $N$-dimensional field space, we use the term vielbein.} 
\begin{equation}
    \label{vielb}
    g_{IJ}(v)  e_i{}^I(v) e_j{}^J(v) \; = \; \delta_{ij} \, ,
\end{equation}
which implies that
\begin{equation}
    \label{vielbreal}
    e_i{}^I(v) \; = \; \frac{\partial \Phi^I}{\partial \Phi'^i} \Big|_v \, , \qquad e_j{}^J(v) \; = \; \frac{\partial \Phi^J}{\partial \Phi'^j} \Big|_v \, .
\end{equation}
We note that the lowercase Latin indices $i,j,k, \ldots,$ correspond to the \textit{mass eigenbasis}, and the indices are raised and lowered with the Kronecker delta $\delta^{ij}$ and $\delta_{ij}$, respectively. The inverse vielbein $e^i{}_I(v)$ is related to the vielbein $e_i{}^I(v)$ via the relation $e^i{}_I = \delta^{ij} e_j{}^J g_{JI}$.

However, the vielbein only flattens the metric and canonically normalizes the linearized equation of motion. To ensure the full transformation from the flavor to the mass eigenbasis, one also needs to diagonalize the mass matrix. This can always be achieved by combining the vielbein field with an orthogonal rotation matrix. Therefore, in the mass eigenbasis the mass matrix becomes
\begin{equation}
    V_{IJ}(v) e_i{}^I(v) e_j{}^J(v) \; = \; V_{ij}(v) \; = \; m_i^2 \delta_{ij} \, .
\end{equation}

\subsubsection{Complex scalar fields}
Following the same steps as before, we consider a theory of $N$ massive complex scalar fields $Z^I$ with general two-derivative interactions and an arbitrary potential $V(Z, \bar{Z})$, given by the Lagrangian
\begin{equation}
    \label{lagcomplex}
    \mathcal{L} \; = \; g_{I \bar{J}}(Z, \bar{Z}) \partial_{\mu} Z^I \partial^{\mu} \bar{Z}^{\bar{J}} - V(Z, \bar{Z}) \, ,
\end{equation}
where $Z^I$ is a complex scalar field and $g_{I \bar{J}}$ is an arbitrary Hermitian metric tensor. As mentioned in the Introduction, we use the convention where the unbarred (holomorphic) indices are on the left, barred (anti-holomorphic) indices are on the right, so that $K^{I \bar{J}}K_{M \bar{J}} = \delta^I_M$. Here we focus on complex K\"ahler manifolds that arise naturally in theories of supersymmetry and supergravity. Note that, a K\"ahler manifold admits a K\"ahler metric which can be understood as a Riemannian metric on a complex manifold that is Hermitian, and the symmetries of K\"ahler metrics are characterized by holomorphic Killing vectors.\footnote{For an interested reader, see Ref.~\cite{Griffithsbook} for an extensive treatment of K\"ahler manifolds, and Ref.~\cite{Freedman:2012zz} for the discussion of complex manifolds in the context of supersymmetry and supergravity.} 

The coordinate transformation formulas for the map $(Z^I, \bar{Z}^{\bar{I}}) \rightarrow (Z'^I, \bar{Z}'^{\bar{I}})$ are given by:
\begin{eqnarray}
    \label{transflawscom}
    V(Z, \bar{Z}) &\rightarrow&  V(Z', \bar{Z}') \, , \\
    \partial_{\mu} Z^I &\rightarrow&  \frac{\partial Z'^I}{\partial Z^J} \partial_{\mu} Z^J \, ,\\
     \partial_{\mu} \bar{Z}^{\bar{I}} &\rightarrow& \frac{\partial \bar{Z}'^{\bar{I}}}{\partial \bar{Z}^{\bar{J}}} \partial_{\mu} \bar{Z}^{\bar{J}} \, , \\
    K_{I\bar{J}} (Z, \bar{Z}) &\rightarrow& \frac{\partial Z^K}{\partial Z'^I} \frac{\partial Z^{\bar{L}}}{\partial Z'^{\bar{J}}} K_{K\bar{L}}(Z', \bar{Z}') \, .
\end{eqnarray}
Since an $N$-dimensional K\"ahler manifold can be interpreted as a $2N$-dimensional real manifold that is parametrized in terms of $N$ complex coordinates, this implies that the physical on-shell scattering amplitudes of the $2N$ real scalar fields remain invariant under the complex coordinate transformations.

One can introduce the complex field expansion
\begin{equation}
    \label{vaccomplex}
    Z^I \; = \; w^I + z^I, \qquad \bar{Z}^{\bar{I}} \; = \; \bar{w}^{\bar{I}} + \bar{z}^{\bar{I}} \, ,
\end{equation}
where $w^I$ is the VEV of the complex scalar field and $z^I$ is the dynamical field fluctuation. Following our discussion for the real scalar fields, we introduce a complex coordinate transformation that canonically normalizes the K\"ahler metric,
\begin{equation}
    K_{I \bar{J}}(w, \bar{w}) \, \frac{\partial Z^I}{\partial Z'^{\alpha}} \Big|_w \frac{\partial \bar{Z}^{\bar{J}}}{\partial \bar{Z}'^{\bar{\beta}}} \Big|_{\bar{w}} \; = \; \delta_{\alpha \bar{\beta}} \, ,
\end{equation}
and similarly, we can express this transformation in terms of complex vielbeins as
\begin{equation}
    \label{comvielbconst}
    K_{I \bar{J}} (w, \bar{w}) e_\alpha{}^I(w,\bar{w}) e_{\bar{\beta}}{}^{\bar{J}}(w,\bar{w}) \; = \; \delta_{\alpha \bar{\beta}} \, .
\end{equation}

When using the complex notation, the Greek indices $\alpha, \beta, \dots$ correspond to the \textit{mass eigenbasis}, and the indices are raised and lowered with the Kronecker delta $\delta^{\alpha \bar{\beta},}$ and $\delta_{\alpha \bar{\beta}}$, respectively. The inverse complex vielbein $e^{\alpha}{}_I(w, \bar{w})$ is related to the complex vielbein via the relation $e^{\alpha}{}_I = \delta^{\alpha \bar{\beta}}e_{\bar{\beta}}{}^{\bar{J}}K_{I \bar{J}}$.

In practice, we want to compute the scattering amplitudes of the real scalar fields in the \textit{mass eigenbasis}, i.e., when the linearized equation of motion is canonically normalized and diagonal. The complex scalar fields can be expressed in terms of their real and imaginary components:
\begin{equation}
    \label{comexp}
    Z^I \; = \; \frac{1}{\sqrt{2}} \left(X^I + i Y^I \right) \, ,
\end{equation}
and the real and imaginary fields can be further expanded as
\begin{equation}
    \label{expandreal}
    X^I(x) \; = \; v^I + \chi^I(x), \qquad Y^I(x) \; = \; u^I + \xi^I(x) \, ,
\end{equation}
where $v^I$ and $u^I$ are the VEVs of the dynamical field fluctuations $\chi^I$ and $\xi^I$. By comparing this expansion to Eq.~(\ref{vaccomplex}), we find that the complex field VEV and fluctuation can be related to the real and imaginary components via
\begin{equation}
    w^I \; = \; \frac{1}{\sqrt{2}} \left(v^I + i u^I \right)\,, \qquad z^I \; = \; \frac{1}{\sqrt{2}} \left(\chi^I + i \xi^I \right) \, .
\end{equation}
If we use the complex field expansion~(\ref{comexp}), the Lagrangian~(\ref{lagcomplex}) becomes~\cite{Ellis:2014opa}:
\begin{equation}
\begin{aligned}
\label{eqgencomp}
\mathcal{L} &\; = \;\frac{1}{2}\left(\partial_\mu X^I, \partial_\mu Y^I\right)\left(\begin{array}{cc}
K_{I \bar{J}}^{\mathcal{R}} & K_{I \bar{J}}^{\mathcal{I}} \\
-K_{I \bar{J}}^{\mathcal{I}} & K_{I \bar{J}}^{\mathcal{R}}
\end{array}\right)\left(\begin{array}{c}
\partial^\mu X^{J} \\
\partial^\mu Y^{J}
\end{array}\right)-V(X, Y) \\
& \; \equiv \; \frac{1}{2} G_{a b} \partial_\mu \Phi^{a} \partial^\mu \Phi^{b}-V(\Phi) \, ,
\end{aligned}
\end{equation}
where $a, b= 1, 2, \ldots, 2N$, and $K_{I \bar{J}}^{\cal{R}}$ and $K_{I \bar{J}}^{\cal{I}}$ denote the real and imaginary parts of the K\"ahler metric, respectively. 

We next introduce the covariant derivatives acting on the  scalar potential $V(Z, \bar{Z})$:
\begin{equation}
    V_{I_1 \ldots I_k}(w, \bar{w}) \; = \; \nabla_{I_1 \ldots} \nabla_{I_k} V(Z, \bar{Z})|_{w, \bar{w}} \, , \quad \partial_I V(Z, \bar{Z})|_{w, \bar{w}}  \; = \; 0 \, ,
\end{equation}
where the covariant derivatives do not need to be symmetrized due to additional symmetries arising in K\"ahler manifolds, and the derivatives can be taken with respect to the fields $Z^I$ or their complex conjugate fields $\bar{Z}^{\bar{I}}$. Here the covariant derivative of a tensor is given by Eq.~(\ref{eq:gentender}). The K\"ahler metric and Hermitian conditions imply that the Christoffel symbols of the form $\Gamma^{I}_{J \bar{K}} = \Gamma^{\bar{I}}_{\bar{J} K}$ vanish, and the only non-vanishing Christoffel symbols are \cite{Wess:1992cp}
\begin{equation}
    \label{eq:complchrist}
    \Gamma^{I}_{J K} \; = \; g^{I \bar{L}} g_{K \bar{L}, J}, \qquad \Gamma^{\bar{I}}_{\bar{J} \bar{K}} \; = \; g^{L \bar{I}} g_{L \bar{K}, \bar{J}} \, .
\end{equation}

The scalar mass matrix is given by the following form
\begin{equation}
    M^2 \; = \; 
    \begin{pmatrix}
        V_{I \bar{J}} & V_{I J} \\
        V_{\bar{I} \bar{J}} & V_{\bar{I} J}
    \end{pmatrix} \, .
\end{equation}

For supersymmetry-preserving vacua, to transform from the \textit{flavor eigenbasis} to the \textit{mass eigenbasis}, we need to rotate the mass matrix with a unitary matrix. Therefore, we can always combine the complex vielbeins with a unitary rotation matrix that will ensure that the above mass matrix is diagonal and transform the Lagrangian~(\ref{lagcomplex}) to the \textit{mass eigenbasis}. In this work, we do not discuss theories involving supersymmetry breaking.

\subsection{Riemann Normal Coordinates}
In this section, we compute the three- and four-point scattering amplitudes by expanding the general two-derivative interaction Lagrangian~(\ref{lag1}) with $N$ massive real scalars $\Phi^I$ and an arbitrary potential $V(\phi)$ in terms of the Riemann normal coordinates.

The general transformation of the original field $\Phi^I = v^I + \phi^I$ in the Lagrangian with metric $g_{IJ}(\phi)$ to the Riemann normal coordinates $\phi^i$ can be obtained from the results in Ref.~\cite{Higashijima:2002fq, Hatzinikitas:2000xe}. They are given by
\begin{equation}
    \label{rnctrans1}
    \phi^I  \longrightarrow \phi^i - \sum_{N=2}^\infty\frac{1}{N!} 
    ~\Gamma^i_{~j_1 j_2\dots j_N}\Big|_{v^i} \, \phi^{j_1}\phi^{j_2}\dots \phi^{j_N}~,
\end{equation}
where the capital letters $I,J, \ldots$ denote the original  basis and the lowercase letters $i, j, \ldots$ denote the Riemann normal coordinate basis. Here the generalized Christoffel symbols $\Gamma^i_{~j_1 j_2\dots j_N}$ are evaluated at the background field value $v^i$ and are defined by
\begin{equation}
    \Gamma^i_{~j_1 j_2\dots j_N} \; = \; \nabla_{j_1} \nabla_{j_2} \ldots \nabla_{j_{N-2}} \Gamma^i{}_{j_{N-1} \, j_N} \, ,
    \label{gengam}
\end{equation}
with $\Gamma^i_{~j_1 j_2}$ the Christoffel symbol defined by Eq.~(\ref{christofsymbols}) obtained from the metric $g_{IJ}(\Phi)$ after applying the vielbeins so that $\Gamma^i_{~j_1 j_2} = e^i_{~I} e_{j_1}^{~J_1} e_{j_2}^{~J_2} \Gamma^I_{~J_1 J_2}$ and similarly for the generalized Christoffel symbols.
In (\ref{gengam}),  the covariant derivatives are only applied on the lower indices following the convention used in Ref.~\cite{Higashijima:2002fq}. Thus, we find
\begin{eqnarray}
    \Gamma^{i}_{~j_1j_2j_3}&=& 
    \Gamma^{i}_{~j_1j_2,\,j_3} -2 \Gamma^i_{~j_1\rho }\Gamma^\rho_{~j_2j_3}   \; = \; \nabla_{j_3} \Gamma^i_{~j_1j_2} \,, \\
    \Gamma^{i}_{~j_1j_2j_3j_4}
    &=& 
    \Gamma^i_{~j_1j_2,\,j_3j_4}-\Gamma^{i}_{~j_1j_2,\,\rho}\Gamma^\rho_{~j_3j_4}-4\Gamma^{i}_{~j_1\rho,\,j_2}\Gamma^\rho_{~j_3j_4}-2\Gamma^\rho_{~j_1j_2,\,j_3}\Gamma^i_{~\rho\,j_4}\nonumber\\
    &&+\,4 \Gamma^i_{~j_1\sigma}\Gamma^\sigma_{~j_2\rho}\Gamma^\rho_{~j_3j_4}+2 \Gamma^\rho_{~j_1 j_2}\Gamma^\sigma_{~j_3j_4}\Gamma^i_{~\rho\sigma}\nonumber\\
    &=&\nabla_{j_4} \nabla_{j_3} \Gamma^i_{~j_1j_2} \,.
\end{eqnarray}

If we use the above transformations~(\ref{rnctrans1}) which diagonalize the metric and then canonically normalize the fields with the additional transformation $\phi^i \rightarrow g^{-1/2}_{\phi^i \phi^i} \phi^i$, which ensures that $g_{ij}(v) = \delta_{ij}(v)$, we find that the general two-derivative interaction Lagrangian~(\ref{lag1}) can be expressed as
\begin{equation}
    \label{lagrnc}
    {\cal{L}} \; = \; \frac{1}{2}\left(\delta_{ij}-\frac{1}{3}R_{ikjl}\phi^k\phi^l
    -\frac{1}{6}\nabla_k R_{iljm}\phi^k\phi^l\phi^m+\dots\right)\partial_\mu\phi^i \partial^\mu\phi^j - V(\phi) \, ,
\end{equation}
where the definition of the Riemann curvature tensor is given by
\begin{equation}
    \label{riemtensor}
    R^I_{~J K L} \; = \; \partial_K \Gamma_{~L J}^I-\partial_L \Gamma_{~K J}^I+\Gamma_{~K M}^I \Gamma_{~L J}^M-\Gamma_{~L M}^I \Gamma_{~K J}^M \, .
\end{equation}
One can also easily obtain the Lagrangian (\ref{lagrnc}) by expanding about the VEV of an arbitrary symmetric metric $g_{IJ}(\Phi)$ in Riemann normal coordinates using the vielbeins~(\ref{vielbreal}). In this case, the scalar fields are given by $\phi^i= e_I{}^i \phi^I $, the arbitrary metric becomes flat, $\delta_{ij} = e_i{}^I e_j{}^J g_{IJ} $, and the Riemann curvature tensor in Riemann normal coordinates is $R_{ijkl}= e_i{}^Ie_j{}^Je_k{}^Ke_l{}^L R_{IJKL}$. We note that in Riemann normal coordinates there is no cubic derivative interaction which greatly simplifies the calculation of scattering amplitudes.

Compact expressions for $\phi^i$ scattering amplitudes can then be written in terms of invariant geometric quantities. Note that the amplitudes should satisfy crossing symmetry for consistency. For example, four-point amplitudes must satisfy 
\begin{equation}
 A_4^{i_1i_2i_3i_4} (s_{12},s_{13},s_{14})
 = A_4^{i_1i_3i_2i_4} (s_{13},s_{12},s_{14})
 = A_4^{i_1i_4i_3i_2} (s_{14},s_{13},s_{12})
\ , \label{eq:crossing}
\end{equation}
where $s_{12} = (p_1+p_2)^2$, $s_{13} = (p_1-p_3)^2$ and $s_{14} = (p_1-p_4)^2$  are the Mandelstam invariants. 
In particular, the expressions for three- and four-point scattering amplitudes are given by
\begin{eqnarray}
    A_3^{i_1i_2i_3}&=&-V^{i_1 i_2 i_3}~, \label{eq:A3amp}\\
    A_4^{i_1i_2i_3i_4}&=&R^{i_1 i_3 i_2i_4}s_{12}+R^{i_1 i_2 i_3i_4}s_{13}-\frac{1}{3}(R^{i_1 i_2 i_3i_4}+R^{i_1 i_3 i_2i_4})(s_{12}+s_{13}+s_{14})~\nonumber\\
    &&-V^{i_1 i_2 i_3i_4}-\sum_j \left( \frac{V^{i_1 i_2 j}V_j^{i_3 i_4}}{s_{12}-m_j^2}+ \frac{V^{i_2 i_3 j}V_j^{i_1 i_4}}{s_{14}-m_j^2}+\frac{V^{i_1 i_3 j}V_j^{i_2 i_4}}{s_{13}-m_j^2} \right)~,
        \label{eq:A4amp}
\end{eqnarray}
where  $V_{I_1 \dots I_k}=\nabla_{(I_1\dots} \nabla_{I_k)} V$, and $V_{i_1 \dots i_k}=e_{i_1}{}^{I_1}\dots e_{i_k}{}^{I_k}V_{I_1 \dots I_k}$.
The expression (\ref{eq:A4amp}), which can be readily checked to satisfy crossing symmetry,  generalizes the kinetic energy part given in Ref.~\cite{Cheung:2021yog} for massless particles to the case of massive particles states. Note that we can also use the transformation of the fields to Riemann normal coordinates to evaluate the potential. In this case, the derivatives in Eqs.~(\ref{eq:A3amp}) and (\ref{eq:A4amp}) are simply partial derivatives. Furthermore, the Riemann tensor (as well as other tensors) with upper small case Latin indices are equal to the Riemann tensor with lower indices since in this basis, indices are raised and lowered with $\delta_{ij}$.

\subsection{K\"ahler Normal Coordinates}
We now generalize the normal coordinate expansion to K\"ahler manifolds and follow the treatment presented in Refs.~\cite{Higashijima:2000wz, Higashijima:2002fq}. When defining the transformations to normal coordinates, one must ensure that the complex structure is always preserved at all orders. 

We first define the K\"ahler potential $K(Z, \bar{Z})$, and the K\"ahler metric is $K_{I \bar{J}}(Z, \bar{Z}) = g_{I \bar{J}}(Z, \bar{Z})$. To find the K\"ahler normal coordinates, we start with the Taylor series expansion of the K\"ahler potential around an arbitrary background field value $Z^I = w^I + z^I$ up to quartic order:
\begin{align}
    \label{kahexp1}
    K(Z, \bar{Z}) & \; = \; \sum_{N,M=0}^{\infty} \frac{1}{N! M!} K_{I_1 \ldots I_N \bar{J}_1 \ldots \bar{J}_{M}} \big|_{w} \, z^{I_1} \ldots z^{I_N} \bar{z}^{\bar{J}_1} \ldots \bar{z}^{\bar{J}_M} ~\nonumber \\
    & \; = \; K(w, \bar{w}) + f(z) + \bar{f}(\bar{z}) + g_{I \bar{J}} \big|_{w} z^I \bar{z}^{\bar{J}} + \frac{1}{2} g_{I \bar{J}, \bar{K}} \big|_{w} z^{I} \bar{z}^{\bar{J}} \bar{z}^{\bar{K}} + \frac{1}{2} g_{J \bar{I}, K} \big|_{w} \bar{z}^{\bar{I}} z^J z^K ~\nonumber \\
    & \; + \; \frac{1}{6} g_{I \bar{J}, \bar{K} \bar{L}}\big|_{w} z^{I} \bar{z}^{\bar{J}} \bar{z}^{\bar{K}} \bar{z}^{\bar{L}} + \frac{1}{4} g_{I \bar{J}, K \bar{L}}\big|_{w} z^I \bar{z}^{\bar{J}} z^K \bar{z}^{\bar{L}} + \frac{1}{6} g_{J \bar{I}, K L} \big|_{w} \bar{z}^{\bar{I}} z^J z^K z^L + \ldots \, ,
\end{align}
where the holomorphic and anti-holomorphic terms have been eliminated using the K\"ahler transformation terms $f(z)$ and $\bar{f}(\bar{z})$.
As shown in Ref.~\cite{Higashijima:2002fq}, one can introduce a holomorphic coordinate transformation of the field fluctuations $z^I$
\begin{align}   
     \label{eq:transkah1}
     z^I \longrightarrow z^{\alpha} - \sum_{N=2}^{\infty} \frac{1}{N!} \Gamma^{\alpha}_{~\beta_1 \beta_2 \ldots \beta_N} \Big|_w z^{\beta_1}z^{\beta_2} \ldots z^{\beta_N} \, ,
\end{align}
where
\begin{equation}
    \Gamma^{\alpha}_{~\beta_1 \beta_2 \ldots \beta_N}  \; = \; \nabla_{\beta_1} \nabla_{\beta_2} \ldots \nabla_{\beta_{N-2}} \Gamma^{\alpha}_{~\beta_{N-1} \beta_N}
\end{equation}
are the covariant derivatives that act only on the lower indices. The capital Latin letters $I,J, \ldots$ denote the original basis and the Greek letters $\alpha, \beta, \ldots$ denote the K\"ahler normal coordinate basis. Importantly, this transformation eliminates terms of the form $z^{\alpha_1} \bar{z}^{\bar{\beta}_1} \ldots \bar{z}^{\bar{\beta}_N}$ and $z^{\alpha_1} \ldots z^{\beta_N} \bar{z}^{\bar{\beta}_1}$ for $N \geq 2$ in the Taylor series. 

The inverse transformation to Eq.~(\ref{eq:transkah1}) is given by
\begin{align}
    \label{eq:invtranskah1}
    z^{\alpha} & \longrightarrow z^I + \sum_{N=2}^{\infty} \frac{1}{N!} g^{I \bar{J}} K_{I_1 \ldots I_N \bar{J}} \big|_w~z^{I_1} \ldots z^{I_N}  \\
    & \; = \; z^I + \sum_{N=2}^{\infty} \frac{1}{N!} g^{I \bar{J}} g_{I_1 \bar{J}, I_2 \ldots I_N} \big|_w~z^{I_1} \ldots z^{I_N} \, .
\end{align}
As shown in~\cite{Higashijima:2000wz, Higashijima:2002fq}, K\"ahler normal coordinates must satisfy the expression
\begin{equation}
    K_{\alpha_1 \ldots \alpha_N \bar{\beta}} (z, \bar{z})\big|_w \; = \; g_{\alpha_1 \bar{\beta}, \alpha_2 \ldots \alpha_N}(z, \bar{z}) \big|_w \; = \; 0 \, ,
\end{equation}
or equivalently,
\begin{equation}
    \partial_{\beta_1} \ldots \partial_{\beta_{N-2}} \Gamma^{\alpha}_{~\beta_{N-1} \beta_N} (z, \bar{z}) \big|_w \; = \; 0 \, .
\end{equation}

Using K\"ahler normal coordinates, the expansion~(\ref{kahexp1}) up to quartic order becomes
\begin{align}
    K(Z, \bar{Z}) \; = \; K(w, \bar{w}) + f(w) + \bar{f}(\bar{w}) + g_{\alpha \bar{\beta}}(w, \bar{w}) z^{\alpha} \bar{z}^{\bar{\beta}} + \frac{1}{4} R_{\alpha \bar{\beta} \gamma \bar{\delta}}(w, \bar{w}) z^{\alpha} \bar{z}^{\bar{\beta}} z^{\gamma} \bar{z}^{\bar{\delta}}     \, ,
\end{align}
where the definition of the K\"ahler curvature tensor is given by \cite{Wess:1992cp}
\begin{equation}
    R_{I\bar{J}K \bar{L}} \; = \; g_{M {\bar{L}}} \Gamma^M{}_{IK,\bar{J}} \; = \; \partial_{K} \partial_{\bar{L}} g_{I \bar{J}}-g^{N \bar{M}} \partial_{K} g_{I \bar{M}} \partial_{\bar{L}} g_{N \bar{J}} \, .
    \label{kahcurvtens}
\end{equation}
The K\"ahler curvature tensor has the following symmetry properties:
\begin{equation}
    \label{eq:kahprop1}
    R_{I \bar{J} K \bar{L}} \; = \; - R_{I \bar{J} \bar{L} K} \; = \; - R_{\bar{J} I  K \bar{L}}\; = \; R_{K \bar{L} I  \bar{J}} \, ,
\end{equation}
and
\begin{equation}
    \label{eq:kahprop2}
    R_{I \bar{J} K \bar{L}} \; = \; R_{K \bar{J} I \bar{L}} \; = \; R_{I \bar{L} K \bar{J}} \, .
\end{equation}

Therefore, one can use the K\"ahler normal coordinate transformations, that diagonalize the K\"ahler metric, and then canonically normalize the complex fields with the additional transformation $z^{\alpha} \rightarrow K^{-1/2}_{z^{\alpha} z^{\alpha}} z^{\alpha}$, which ensures that $K_{\alpha \bar{\beta}}(w, \bar{w}) = \delta_{\alpha \bar{\beta}}(w, \bar{w})$. Thus, in K\"ahler normal coordinates, the general two-derivative interaction Lagrangian with $N$ massive complex scalar fields~(\ref{lagcomplex}) becomes
\begin{equation}
    \label{eq:KNClagrangian}
    \mathcal{L} \; = \; \left( \delta_{\alpha \bar{\beta}} + R_{\alpha \bar{\beta} \gamma \bar{\delta}} z^{\gamma} \bar{z}^{\bar{\delta}}  + \ldots \right) \partial_{\mu} z^{\alpha} \partial^{\mu} \bar{z}^{\bar{\beta}} - V(z, \bar{z}) \, .
\end{equation}
This Lagrangian can also be found by expanding about the VEV of a symmetric K\"ahler metric $K_{I \bar{J}}$ in K\"ahler normal coordinates and using the complex vielbeins. Then, the complex scalar fields can be expressed as $z^{\alpha} = e_I{}^{\alpha} z^I$ and $\bar{z}^{\bar{\alpha}} = \bar{e}_{\bar{I}}{}^{\bar{\alpha}} \bar{z}^{\bar{I}}$, the K\"ahler metric becomes flat, $\delta_{\alpha \bar{\beta}} = e_{\alpha}{}^I \bar{e}_{\bar{\beta}}{}^{\bar{J}}K_{I \bar{J}}$, and the K\"ahler curvature tensor in K\"ahler normal coordinates is given by $R_{\alpha \bar{\beta} \gamma \bar{\delta}} = e_{\alpha}{}^{I} \bar{e}_{\bar{\beta}}{}^{\bar{J}} e_{\gamma}{}^{K} \bar{e}_{\bar{\delta}}{}^{\bar{L}} R_{I \bar{J} K \bar{L}}$. As in Riemann normal coordinates, there are no cubic derivative interactions in K\"ahler normal coordinates and the scattering amplitude computation is significantly simplified.

To compute the scalar field interactions from the Lagrangian~(\ref{eq:KNClagrangian}), it is convenient to split the Lagrangian into kinetic and potential interaction terms. The kinetic terms in K\"ahler normal coordinates up to quartic order are given by
\begin{equation}
     \label{eq:kintermsKNC}
        \mathcal{K}_{\rm{KNC}} \; = \; \partial_{\mu} z^{\alpha} \partial^{\mu} \bar{z}_{\alpha} + R_{\alpha \bar{\beta} \gamma \bar{\delta}} z^{\gamma} \bar{z}^{\bar{\delta}}  \partial_{\mu} z^{\alpha} \partial^{\mu} \bar{z}^{\bar{\beta}} +\dots\, ,
\end{equation}
where in terms of the K\"ahler potential, the K\"ahler curvature tensor $R_{\alpha \bar{\beta} \gamma \bar{\delta}}=K_{\alpha \bar{\beta},\gamma \bar{\delta}}-K^{\mu\bar{\nu}} K_{\mu\bar{\beta},\bar{\delta}}K_{\alpha\bar{\nu},\gamma}$ involves just ordinary partial derivatives.
In this case, the use of complex field notation is
the most compact form of writing scattering amplitudes. Indeed, from \eqref{eq:kintermsKNC} we see that only a four-point amplitude involving two complex and two complex conjugate fields is nonzero. One finds the simple expression
\begin{equation}
A_{4,kin}^{z_{\alpha_1} z_{\alpha_2} {\bar z}_{\bar \alpha_3} {\bar z}_{\bar \alpha_4} } \; = \; R^{\alpha_1 \bar{\alpha}_3 \alpha_2 \bar{\alpha}_4} s_{12}  \ , 
\label{eq:complex1} 
\end{equation}      
whereas the four-point amplitudes with permuted indices are easily obtained by crossing symmetry
\begin{equation}
A_{4,kin}^{z_{\alpha_1} {\bar z}_{\bar \alpha_2} z_{\alpha_3}  {\bar z}_{\bar \alpha_4} } \; = \; R^{\alpha_1 \bar{\alpha}_2 \alpha_3 \bar{\alpha}_4} s_{13}  \ , \ \qquad A_{4,kin}^{z_{\alpha_1} {\bar z}_{\bar \alpha_2} {\bar z}_{\bar \alpha_3} 
 z_{\alpha_4}   } \; = \; R^{\alpha_1 \bar{\alpha}_2 \alpha_4 \bar{\alpha}_3} s_{14} \ . 
\label{eq:complex2} 
\end{equation}
One can also obtain the four-point amplitudes for real fields.
If we use the field expansion $z^{\alpha} = \frac{1}{\sqrt{2}} \left(\chi^{\alpha} + i \xi^{\alpha} \right)$ and $\bar{z}^{\bar{\alpha}} = \frac{1}{\sqrt{2}} \left(\chi^{\bar{\alpha}} - i \xi^{\bar{\alpha}} \right) = \bar{z}^{\alpha} = \frac{1}{\sqrt{2}} \left(\chi^{\alpha} - i \xi^{\alpha} \right)$, the kinetic terms of the Lagrangian become
\begin{eqnarray}
    \mathcal{K}_{\rm{KNC}} & = & \frac{1}{2}\partial_{\mu} \chi^{\alpha} \partial^{\mu}\chi_{\alpha} + \frac{1}{2}\partial_{\mu} \xi^{\alpha} \partial^{\mu}\xi_{\alpha} \nonumber\\
    &&+ \frac{1}{4}R_{\alpha \bar{\beta} \gamma \bar{\delta}} \left[ ( \partial_{\mu} \chi^{\alpha} \partial^{\mu} \chi^{\bar{\beta}}  + \partial_{\mu} \xi^{\alpha} \partial^{\mu} \xi^{\bar{\beta}}) (\chi^{\gamma} \chi^{\bar{\delta}} +  \xi^{\gamma} \xi^{\bar{\delta}} -i (\chi^{\gamma} \xi^{\bar{\delta}} -  \xi^{\gamma} \chi^{\bar{\delta}}))\right. \nonumber \\
    && \left. - ( \partial_{\mu} \chi^{\alpha} \partial^{\mu} \xi^{\bar{\beta}}  
    - \partial_{\mu} \xi^{\alpha} \partial^{\mu} \chi^{\bar{\beta}}) (\chi^{\gamma} \xi^{\bar{\delta}} -  \xi^{\gamma} \chi^{\bar{\delta}} +i (\chi^{\gamma} \chi^{\bar{\delta}} +  \xi^{\gamma} \xi^{\bar{\delta}})) \right]~.
    \end{eqnarray}
Using the above kinetic terms of the Lagrangian, we find that the four-point scattering amplitudes are given by 
\begin{eqnarray}
    A_{4, \, kin}^{\chi_{\alpha_1}\chi_{\alpha_2}\chi_{\alpha_3}\chi_{\alpha_4}}&=&  A_{4, \, kin}^{\xi_{\alpha_1}\xi_{\alpha_2}\xi_{\alpha_3}\xi_{\alpha_4}} = \frac{1}{4}\left(R^{\alpha_1 \bar{\alpha}_3 \alpha_2 \bar{\alpha}_4} s_{12} + R^{\alpha_1 \bar{\alpha}_2 \alpha_3 \bar{\alpha}_4} s_{13} + R^{\alpha_1 \bar{\alpha}_2 \alpha_4 \bar{\alpha}_3} s_{14}\right) + \rm{c.c.}~\nonumber , \\
    A_{4, \, kin}^{\chi_{\alpha_1}\chi_{\alpha_2}\xi_{\alpha_3}\xi_{\alpha_4}}&=& A_{4, \, kin}^{\xi_{\alpha_1}\xi_{\alpha_2}\chi_{\alpha_3}\chi_{\alpha_4}} = \frac{1}{4}\left(-R^{\alpha_1 \bar{\alpha}_3 \alpha_2 \bar{\alpha}_4} s_{12} + R^{\alpha_1 \bar{\alpha}_2 \alpha_3 \bar{\alpha}_4} s_{13} + R^{\alpha_1 \bar{\alpha}_2 \alpha_4 \bar{\alpha}_3} s_{14}\right) + \rm{c.c.} ~\nonumber , \\
    A_{4, \, kin}^{\chi_{\alpha_1}\xi_{\alpha_2}\chi_{\alpha_3}\xi_{\alpha_4}}&=& A_{4, \, kin}^{\xi_{\alpha_1}\chi_{\alpha_2}\xi_{\alpha_3}\chi_{\alpha_4}} = \frac{1}{4}\left(R^{\alpha_1 \bar{\alpha}_3 \alpha_2 \bar{\alpha}_4} s_{12} -  R^{\alpha_1 \bar{\alpha}_2 \alpha_3 \bar{\alpha}_4} s_{13} + R^{\alpha_1 \bar{\alpha}_2 \alpha_4 \bar{\alpha}_3} s_{14}\right) + \rm{c.c.} ~\nonumber , \\
     A_{4, \, kin}^{\chi_{\alpha_1}\xi_{\alpha_2}\xi_{\alpha_3}\chi_{\alpha_4}}&=& A_{4, \, kin}^{\xi_{\alpha_1}\chi_{\alpha_2}\chi_{\alpha_3}\xi_{\alpha_4}} = \frac{1}{4}\left(R^{\alpha_1 \bar{\alpha}_3 \alpha_2 \bar{\alpha}_4} s_{12} + R^{\alpha_1 \bar{\alpha}_2 \alpha_3 \bar{\alpha}_4} s_{13} - R^{\alpha_1 \bar{\alpha}_2 \alpha_4 \bar{\alpha}_3} s_{14}\right) + \rm{c.c.} \, , \nonumber\\
      A_{4, \, kin}^{\chi_{\alpha_1}\chi_{\alpha_2}\chi_{\alpha_3}\xi_{\alpha_4}}&=& - A_{4, \, kin}^{\xi_{\alpha_1}\xi_{\alpha_2}\xi_{\alpha_3}\chi_{\alpha_4}} = \frac{i}{4}\left(R^{\alpha_1 \bar{\alpha}_3 \alpha_2 \bar{\alpha}_4} s_{12} + R^{\alpha_1 \bar{\alpha}_2 \alpha_3 \bar{\alpha}_4} s_{13} - R^{\alpha_1 \bar{\alpha}_2 \alpha_4 \bar{\alpha}_3} s_{14}\right) + \rm{c.c.} \, , \nonumber\\
     \label{Amp4kin:KNC}
\end{eqnarray}
which are, of course, equivalent to the more compact formula (\ref{eq:complex1}).  
Since these amplitudes are expressed in terms of the Mandelstam invariants, they are also valid for the massive case.
Note that these expressions are consistent with the complex amplitudes given in Eqs.~(\ref{eq:complex1}) and (\ref{eq:complex2}) as well as crossing symmetry. 

Next, we compute the interactions arising from the following complex Taylor series of the scalar potential up to quartic order \cite{Clark:1987ct}:
\begin{align}
    &V(Z, \bar{Z}) \;  = \;  V(w, \bar{w}) + V_{\alpha} z^{\alpha} + V_{\bar{\alpha}} \bar{z}^{\bar{\alpha}} + \frac{1}{2}V_{\alpha \beta} z^{\alpha} z^{\beta} + V_{\bar{\alpha} \beta} \bar{z}^{\bar{\alpha}} z^{\beta}  + \frac{1}{2}V_{\bar{\alpha} \bar{\beta}} \bar{z}^{\bar{\alpha}} \bar{z}^{\bar{\beta}}+ \frac{1}{6} V_{\alpha \beta \gamma} z^{\alpha} z^{\beta} z^{\gamma}~\nonumber \\ 
    &  + \frac{1}{2} V_{\alpha \beta \bar{\gamma}} z^{\alpha} z^{\beta} \bar{z}^{\bar{\gamma}} + \frac{1}{2} V_{\bar{\alpha} \bar{\beta} \gamma} \bar{z}^{\bar{\alpha}} \bar{z}^{\bar{\beta}} z^{\gamma} + \frac{1}{6} V_{\bar{\alpha} \bar{\beta} \bar{\gamma}} \bar{z}^{\bar{\alpha}} \bar{z}^{\bar{\beta}} \bar{z}^{\bar{\gamma}} + \frac{1}{24} V_{\alpha \beta \gamma \delta} z^{\alpha} z^{\beta} z^{\gamma} z^{\delta} + \frac{1}{6} V_{\alpha \beta \gamma \bar{\delta}}  z^{\alpha} z^{\beta} z^{\gamma} \bar{z}^{\bar{\delta}}~\nonumber \\
    & + \frac{1}{4} \left(V_{\alpha \bar{\gamma} \beta \bar{\delta}} + V_{\bar{\gamma} \alpha \bar{\delta} \beta}
    -\frac{1}{2} (V_{\alpha \beta \bar{\gamma} \bar{\delta}} +V_{\bar{\gamma} \bar{\delta}\alpha \beta}) \right) z^{\alpha} z^{\beta} \bar{z}^{\bar{\gamma}} \bar{z}^{\bar{\delta}} 
    + \frac{1}{6} V_{\alpha \bar{\beta} \bar{\gamma} \bar{\delta}} z^{\alpha} \bar{z}^{\bar{\beta}} \bar{z}^{\bar{\gamma}} \bar{z}^{\bar{\delta}} +\frac{1}{24} V_{\bar{\alpha} \bar{\beta} \bar{\gamma} \bar{\delta}} \bar{z}^{\bar{\alpha}} \bar{z}^{\bar{\beta}} \bar{z}^{\bar{\gamma}} \bar{z}^{\bar{\delta}} \, ,
    \label{Vknctaylor}
\end{align}
where $V_{I_1 \dots I_k}=\nabla_{I_1\dots} \nabla_{I_k} V$ and $V_{\alpha_1 \dots \alpha_k}=e_{\alpha_1}{}^{I_1}\dots e_{\alpha_k}{}^{I_k}V_{I_1 \dots I_k}$.\footnote{We remind the reader that here we consider theories with supersymmetry-preserving vacua.} In the flat limit, (\ref{Vknctaylor}) reduces to the ordinary Taylor expansion.
Note that in this case, the indices are not symmetrized. Using this expansion, one can find the following four-point amplitudes involving complex fields
\begin{align}
     A_{4, \, pot}^{z_{\alpha_1} z_{\alpha_2} z_{\alpha_3} z_{\alpha_4}} \; = \;
     & -V^{\alpha_1 \alpha_2 \alpha_3 \alpha_4}  \nonumber \\
    -\sum_\beta & \left( \frac{V^{\alpha_1 \alpha_2 \rho_\beta}V_{\rho_\beta}{}^{\alpha_3 \alpha_4}}{s_{12}-m_{\beta\bar{\beta}}^2}+ \frac{V^{\alpha_2 \alpha_3 \rho_\beta}V_{\rho_\beta}{}^{\alpha_1 \alpha_4}}{s_{14}-m_{\beta \bar{\beta}}^2}+\frac{V^{\alpha_1 \alpha_3 \rho_\beta}V_{\rho_\beta}{}^{\alpha_2 \alpha_4}}{s_{13}-m_{\beta\bar{\beta}}^2} \right) \, ,
\end{align}
\begin{align}
    A_{4, pot}^{z_{\alpha_1} z_{\alpha_2} \bar{z}_{\bar{\alpha}_3} \bar{z}_{\bar{\alpha}_4}} \; = \;
     & - \left(V^{\alpha_1 \bar{\alpha}_3 \alpha_2 \bar{\alpha}_4} + V^{\bar{\alpha}_3 \alpha_1 \bar{\alpha}_4 \alpha_2} - \frac{1}{2} (V^{\alpha_1 \alpha_2 \bar{\alpha}_3 \bar{\alpha}_4} + V^{\bar{\alpha}_3 \bar{\alpha}_4 \alpha_1 \alpha_2}) \right) \nonumber \\
    -\sum_\beta & \left( \frac{V^{\alpha_1 \alpha_2 \rho_\beta}V_{\rho_\beta}{}^{\bar{\alpha}_3 \bar{\alpha}_4}}{s_{12}-m_{\beta\bar{\beta}}^2}+ \frac{V^{\alpha_2 \bar{\alpha}_3 \rho_\beta}V_{\rho_\beta}{}^{\alpha_1 \bar{\alpha}_4}}{s_{14}-m_{\beta\bar{\beta}}^2}+\frac{V^{\alpha_1 \bar{\alpha}_3 \rho_\beta}V_{\rho_\beta}{}^{\alpha_2 \bar{\alpha}_4}}{s_{13}-m_{\beta \bar{\beta}}^2} \right) \, ,
\end{align}
where here the indices are complex $\rho_{\beta} = \beta, {\bar \beta}$. Note that the written expressions for the contractions over $\rho_\beta$ are shorthand for example,  $V^{\alpha_1 \alpha_2 \rho_\beta}V_{\rho_\beta}{}^{\alpha_3 \alpha_4} = V^{\alpha_1 \alpha_2 \bar\beta}V_{\beta}{}^{\alpha_3 \alpha_4} + V^{\alpha_1 \alpha_2 \beta}V_{\bar\beta}{}^{\alpha_3 \alpha_4}$. Recall also that in this basis, the mass matrix is diagonal and $V_{\alpha \beta} = V_{\bar{\alpha} \bar{\beta}} = 0$ and that indices in this basis can be raised and lowered with a Kronecker delta. We do not list all possible channels that could be readily found from Eq.~(\ref{Vknctaylor}).

One can also find the expressions for the three- and four-point scattering amplitudes for the real fields, similar to Eq.~(\ref{eq:A4amp}). 
The scalar potential gives rise to the following three-point contributions:
\begin{eqnarray}
    V^{\varphi_{\alpha_1} \varphi_{\alpha_2} \varphi_{\alpha_3}} &=& c_1 V_{{\alpha_{(1}} {\alpha_2} {\alpha_{3)}}}  + c_2 V_{\alpha_{(1} \alpha_{2)} \bar{\alpha}_3 }+ c_3 V_{\alpha_{(1} \alpha_{3)} \bar{\alpha}_2 } + c_4 V_{\alpha_{(2} \alpha_{3)} \bar{\alpha}_1 }+ \rm{c.c.} \, ,
    \label{KNC3}
    \end{eqnarray}
    where the real fields are denoted $\varphi = \{\chi, \xi \}$, $c_1 = \frac{1}{2\sqrt{2}}(1, i, -1, -i)$, $c_2 = \frac{1}{2\sqrt{2}}(1, -i, 1, i)$, $c_3 = \frac{1}{2\sqrt{2}}(1, i, 1, i)$, and $c_4 = \frac{1}{2\sqrt{2}}(1, i, -1, i)$, when there are $(0,1,2,3)$ $\xi$ fields among the $\varphi_{\alpha_1},\varphi_{\alpha_2},\varphi_{\alpha_3}$.  
Symmetrized indices in Eq.~(\ref{KNC3}) are enclosed by parentheses.
The four-point contributions are:
    \begin{align}
    V^{\varphi_{\alpha_1} \varphi_{\alpha_2} \varphi_{\alpha_3} \varphi_{\alpha_4}} &=  c_5 V_{\alpha_{(1} \alpha_2 \alpha_3 \alpha_{4)}} + c_6 V_{\alpha_{(1} \alpha_2 \alpha_{3)} \bar{\alpha}_{4}} + c_7 V_{\alpha_{(1} \alpha_2 \alpha_{4)} \bar{\alpha}_{3}} + c_8 V_{\alpha_{(1} \alpha_3 \alpha_{4)} \bar{\alpha}_{2}}
   + c_9 V_{\alpha_{(2} \alpha_3 {\alpha}_{4)} \bar{\alpha}_{1}} ~\nonumber \\ 
    &+c_{10} V_{\alpha_{(1} \alpha_{2)} \bar{\alpha}_{(3} \bar{\alpha}_{4)}} + c_{11} V_{\alpha_{(1} \alpha_{3)} \bar{\alpha}_{(2} \bar{\alpha}_{4)}} 
    + c_{12} V_{\alpha_{(1} \alpha_{4)} \bar{\alpha}_{(2} \bar{\alpha}_{3)}}
    + c_{13} V_{\alpha_{(2} \alpha_{3)} \bar{\alpha}_{(1} \bar{\alpha}_{4)}} ~\nonumber \\ 
   &+ c_{14} V_{\alpha_{(2} \alpha_{4)} \bar{\alpha}_{(1} \bar{\alpha}_{3)}}
    + c_{15} V_{\alpha_{(3} \alpha_{4)} \bar{\alpha}_{(1} \bar{\alpha}_{2)}}
    + c_{16} V_{\alpha_{(1} \bar{\alpha}_{2)} \alpha_{(3} \bar{\alpha}_{4)}} + c_{17} V_{\alpha_{(1} \bar{\alpha}_{3)} \alpha_{(2} \bar{\alpha}_{4)}} ~\nonumber \\
    &+ c_{18} V_{\alpha_{(1} \bar{\alpha}_{4)} \alpha_{(2} \bar{\alpha}_{3)}}  
    + c_{19} V_{\alpha_{(2} \bar{\alpha}_{3)} \alpha_{(1} \bar{\alpha}_{4)}} + c_{20} V_{\alpha_{(2} \bar{\alpha}_{4)} \alpha_{(1} \bar{\alpha}_{3)}} +
    c_{21} V_{\alpha_{(3} \bar{\alpha}_{4)} \alpha_{(1} \bar{\alpha}_{2)}} ~\nonumber \\
    &+ c_{22} V_{\alpha_{(1} \bar{\alpha}_{|(3} \alpha_{4)|} \bar{\alpha}_{2)}} +  
    c_{23} V_{\alpha_{(3} \bar{\alpha}_{|(1} \alpha_{2)|} \bar{\alpha}_{4)}} + \rm{c.c.}
    \, ,
    \label{KNC4}
    \end{align}
where $c_5 = \frac{1}{4}(1, i, -1, -i, 1)$, $c_6 = \frac{1}{4}(1, -i, -1, -i, -1)$, $c_7 = \frac{1}{4}(1, i, -1, i, -1)$, \\ 
$c_8 = \frac{1}{4}(1, i, 1, i, -1)$, $c_9 = \frac{1}{4}(1, i, 1, i, -1)$, $c_{10} = \frac{1}{8} (-1, i, 1, -i, -1)$, $c_{11} = \frac{1}{8} (-1, i, -1, -i, -1)$, $c_{12} = \frac{1}{8}(-1, -i, -1, i, -1)$, $c_{13} = \frac{1}{8}(-1, i, -1, -i, -1)$, $c_{14}=\frac{1}{8}(-1, -i, -1, i, -1)$, \\ $c_{15} = \frac{1}{8} (-1, -i, 1, i, -1)$, $c_{16} = \frac{1}{4} (1, -i, 1, i, 1)$, $c_{17} = \frac{1}{4}(1, -i, -1, i, 1)$, $c_{18} = \frac{1}{4}(1, i, -1, -i, 1)$, $c_{19} = \frac{1}{4}(1, -i, -1, i, 1)$, $c_{20} = \frac{1}{4} (1, i, -1, -i, 1)$, $c_{21} = \frac{1}{4} (1, i, 1, -i, 1)$, $c_{22} = c_{23} = (0, 0, 2, 0, 0)$, when there are $(0,1,2,3,4)$ $\xi$ fields among the four $\varphi_{\alpha_i}$. Again, the symmetrization is over indices within parentheses. The vertical bars delineate separate symmetrizations. 
Note that these coefficients are not unique and a rearrangement of indices leading to different coefficients is possible.
The potential contributions to the three- and four-point amplitudes are then given by
      \begin{eqnarray}
      \label{Amp3:KNC}
    A_{3, \, pot}^{\varphi_{\alpha_1} \varphi_{\alpha_2} \varphi_{\alpha_3}} &=&
    -V^{\varphi_{\alpha_1} \varphi_{\alpha_2} \varphi_{\alpha_3}}\,, \\
      \label{Amp4:KNC}
    A_{4, \, pot}^{\varphi_{\alpha_1} \varphi_{\alpha_2} \varphi_{\alpha_3} \varphi_{\alpha_4}} &=&
    -V^{\varphi_{\alpha_1} \varphi_{\alpha_2} \varphi_{\alpha_3} \varphi_{\alpha_4}}~\nonumber \\
    &-&\sum_\beta \left( \frac{V^{\varphi_{\alpha_1} \varphi_{\alpha_2} \varphi_\beta}V_{\varphi_\beta}{}^{\varphi_{\alpha_3} \varphi_{\alpha_4}}}{s_{12}-m_{\varphi_\beta}^2}+ \frac{V^{\varphi_{\alpha_2} \varphi_{\alpha_3} \varphi_\beta}V_{\varphi_\beta}{}^{\varphi_{\alpha_1} \varphi_{\alpha_4}}}{s_{14}-m_{\varphi_\beta}^2}+\frac{V^{\varphi_{\alpha_1} \varphi_{\alpha_3} \varphi_\beta}V_{\varphi_\beta}{}^{\varphi_{\alpha_2} \varphi_{\alpha_4}}}{s_{13}-m_{\varphi_\beta}^2} \right). ~\nonumber  \\
\end{eqnarray}
The total three-point scattering amplitude only receives a contribution from the potential \eqref{Amp3:KNC}, while the total four-point scattering amplitude is $A_{4}^{\varphi_{\alpha_1} \varphi_{\alpha_2} \varphi_{\alpha_3} \varphi_{\alpha_4}} = A_{4, \, kin}^{\varphi_{\alpha_1} \varphi_{\alpha_2} \varphi_{\alpha_3} \varphi_{\alpha_4}} + A_{4, \, pot}^{\varphi_{\alpha_1} \varphi_{\alpha_2} \varphi_{\alpha_3} \varphi_{\alpha_4}}$ obtained from \eqref{Amp4kin:KNC} and \eqref{Amp4:KNC}.
As in the case of RNC we can also use the transformation of the fields to K\"ahler normal coordinates to evaluate the potential. In this case, the derivatives in Eqs.~(\ref{Amp4:KNC}) are simply partial derivatives.
Crossing symmetry is guaranteed by the symmetrization in Eqs.~(\ref{KNC3}) and (\ref{KNC4}).
\section{Supergravity Framework}
\label{sugr}
We recall that a generic theory of minimal $\mathcal{N} = 1$ supergravity can be characterized by a K\"ahler potential $K(Z, \bar{Z})$, where $Z$ and $\bar{Z}$ are complex scalar fields and their Hermitian conjugates, respectively. To include the field interactions arising from the scalar potential, one can introduce a holomorphic function of the complex fields, $Z$, called the superpotential, $W(Z)$. One may combine the K\"ahler potential $K(Z, \bar{Z})$ and the superpotential $W(Z)$ into the following K\"ahler extended function
\begin{equation}
    G \equiv K + \ln|W|^2 \, .
\end{equation}
Therefore, the bosonic part of the $\mathcal{N} = 1$ supergravity Lagrangian can be written in the form
\begin{equation}
    \label{lagsugra}
    \mathcal{L} \; = \; K_{I \bar{J}}(Z, \bar{Z}) \partial_{\mu} Z^I \partial^{\mu} \bar{Z}^{\bar{J}} - V(Z, \bar{Z}) \, ,
\end{equation}
where the scalar potential can be expressed as
\begin{equation}
    \label{eq:effpotsugra}
    V(Z, \bar{Z}) \; = \; e^G \left[G_I G_{\bar{J}} K^{I \bar{J}}  - 3 \right] \; = \; K_{I \bar{J}} F^I F^{\bar{J}} - 3e^G \, .
\end{equation}
Here $K^{I \bar{J}}$ is the inverse K\"ahler metric, and the $F$-term is given by
\begin{equation}
    F^I \; = \; - e^{G/2} G^{I \bar{J}} G_{\bar{J}} \, ,
\end{equation}
where $G^{I \bar{J}} = K^{I \bar{J}}$. Note that the general $\mathcal{N} = 1$ supergravity Lagrangian (\ref{lagsugra}) coincides with the Lagrangian~(\ref{lagcomplex}), with $g_{I \bar{J}} = K_{I \bar{J}}$. Here we are only considering the $F$-term contribution in the supergravity Lagrangian but in general, one can expect $D$-term contributions for gauge non-singlet fields. The $F$-term contribution characterizes the order of supersymmetry breaking in theories of supergravity once local supersymmetry is broken and the $F$-term obtains a VEV. For minimal supergravity, with $g_{I \bar{J}} = \delta_{I \bar{J}}$ we have $\langle F^I \rangle = \langle e^{K/2}(W^I + K^I W) \rangle = -m_{3/2} \langle G^{I \bar{J}} G_J \rangle  \neq 0$, where the gravitino mass is given by
\begin{equation}
    m_{3/2}^2 \; = \; e^{\langle G \rangle} \, .
\end{equation}

Clearly in minimal supergravity, the above discussion of Riemann and K\"ahler normal coordinates is irrelevant and trivial. As the curvature vanishes,
the calculations of amplitudes is straightforward using only ordinary partial derivatives of the potential. In no-scale supergravity, on the other hand, the curvature is non-trivial. As we have already seen, the K\"ahler curvature tensor is given in Eq.~(\ref{kahcurvtens}). This can be contracted to obtain the Ricci tensor \cite{Grisaru:1982sr}
 \beq
 \label{riccitensor}
 R_{I \bar{J}} = g^{K {\bar L}} R_{I \bar{J} K \bar{L}} = \delta^K{}_M \Gamma^M{}_{I K ,\bar{J}} = \Gamma^K_{I K,\bar{J}} =\left( \ln (-g) \right)_{,I, {\bar J}} \, ,
 \eeq
where $g \equiv \det g_{K \bar{L}}$.

In no-scale supergravity models, with the K\"ahler potential of the form~(\ref{v0}), the logarithm of the determinant of the metric is given by
\beq
\ln(-g) = \ln 3 + N \ln\alpha + \frac{N+1}{3 \alpha} K \, ,
\eeq
where $N$ is the total number of superfields, i.e., $T$ + $N-1$ matter fields $\Phi^I$.
Then the Ricci tensor~(\ref{riccitensor}) is 
\beq
R_{I \bar{J}} \; = \; \frac{N+1}{3 \alpha} g_{I \bar{J}} \, ,
\eeq
and the Ricci scalar is given by
\beq
R \; = \; g^{I \bar{J}} R_{I \bar{J}} = \frac{N(N+1)}{3\alpha} \, ,
\eeq
showing clearly the maximally symmetric property of no-scale models. 

\section{Single Complex Scalar Field Interactions}
\label{su11}
In this section, we examine general non-linear sigma models. We consider two-derivative interaction theories consisting of a single complex scalar field or two real scalar fields. First, we study two-derivative interactions with no potential, arising from a single complex field no-scale supergravity framework, and then compute the scattering amplitudes for different potentials. 

We briefly summarize our notation. When using general vectors and tensors, we use the capital Latin indices $I, J, \ldots \, , $ to denote the original \textit{flavor eigenbasis}. The \textit{mass eigenbasis} in Riemann normal coordinates (RNC) is denoted by the lowercase Latin indices $i, j, \ldots \,$, and the \textit{mass eigenbasis} in K\"ahler normal coordinates (KNC) is denoted by the Greek indices $\alpha, \beta, \ldots , \,$. When we specify the fields, the tilded coordinates correspond to the RNC and the hatted coordinates correspond to KNC.

\subsection{Two-Derivative Interactions with No Potential}
Using the general supergravity Lagrangian (\ref{lagsugra}) together with the K\"ahler potential~(\ref{kah1}), we obtain the following Lagrangian
\begin{equation}
\label{eq:TLag1}
    {\cal L} = \frac{3\alpha}{(T+\bar T)^2} \partial_\mu T \partial^\mu {\bar T} \; = \;  \frac{3\alpha}{4 X^2} \left((\partial_\mu X)^2+(\partial_\mu Y)^2\right)
    \, ,
\end{equation}
assuming the complex field decomposition $T = \frac{1}{\sqrt{2}} \left(X + i Y\right)$. We first focus on the right-hand side of the Lagrangian and discuss the field transformation to RNC and KNC.

Using Eq.~(\ref{expandreal}), one can expand the real scalar fields $X$ and $Y$ in terms of their corresponding VEVs and field fluctuations. Therefore, one can readily transform the right-hand side of the Lagrangian~(\ref{eq:TLag1}) to Riemann normal coordinates by using the following metric
\begin{equation}
g_{I J}(X) \; = \; \frac{3 \alpha}{2X^2}
\begin{pmatrix}
    1 & 0 \\
    0 & 1
    \end{pmatrix} \, ,
\end{equation}
which implies that the vielbein is 
\begin{equation}
    \label{vielb1}
e_i{}^I(v, u) \; = \; \sqrt{\frac{2}{3 \alpha}}v \begin{pmatrix}
    1 & 0 \\
    0 & 1
    \end{pmatrix} \, ,
\end{equation}
ensuring that the vielbein constraint~(\ref{vielb}) is satisfied and the metric evaluated at the VEV is flat. Note that the vielbein only depends on the VEV $v$. Using the definition~(\ref{riemtensor}) and computing the Riemann curvature tensor at $X=v$ , we find that the non-zero components are given by 
\begin{equation}
    R_{\chi \xi \chi \xi} \; = \; R_{\xi \chi \xi \chi} \; = \; -\frac{3 \alpha}{2 v^4} \, , \qquad R_{\chi \xi \xi \chi} \; = \; R_{\xi \chi \chi \xi} \; = \; \frac{3 \alpha}{2 v^4} \, .
\end{equation}
If we use the vielbein~(\ref{vielb1}) together with the identity $R_{ijkl}= e_i{}^Ie_j{}^Je_k{}^Ke_l{}^L R_{IJKL}$, the Riemann curvature tensor in the \textit{mass eigenbasis} becomes 
\begin{equation}
    R_{\tilde{\chi} \tilde{\xi} \tilde{\chi} \tilde{\xi}} \; = \; R_{\tilde{\xi} \tilde{\chi} \tilde{\xi} \tilde{\chi}} \; = \; -\frac{2}{3\alpha} \, , \qquad R_{\tilde{\chi} \tilde{\xi} \tilde{\xi} \tilde{\chi}} \; = \; R_{\tilde{\xi} \tilde{\chi} \tilde{\chi} \tilde{\xi}} \; = \; \frac{2}{3\alpha} \, ,
\end{equation}
where the tilde denotes the RNC basis. Note that in the mass eigenbasis there is no dependence on the VEV.

Using the general two-derivative interaction Lagrangian expansion in the mass eigenbasis~(\ref{lagrnc}), we find that the kinetic terms~(\ref{eq:TLag1}) up to quartic interactions are
\begin{equation}
    \label{eq:RNLag2field}
    \mathcal{K}_{\rm RNC} = \frac{1}{2}(\partial_\mu\tilde \chi)^2+\frac{1}{2}(\partial_\mu\tilde \xi)^2 +\frac{1}{9\alpha}\left({\tilde \chi}\partial_\mu \tilde \xi-{\tilde \xi}\partial_\mu \tilde \chi\right)^2\,.
\end{equation}

Similarly, instead of using the vielbeins, we can transform the fields $(\chi, \xi)$ to Riemann normal coordinates $(\tilde{\chi}, \tilde{\xi})$ via the relations~(\ref{rnctrans1}). These transformations ensure that the metric is diagonal. If we then rescale the fields $\tilde \chi \rightarrow \tilde{g}_{\tilde{\chi} \tilde{\chi}}^{-1/2} \tilde{\chi}$ and $\tilde \xi \rightarrow \tilde{g}_{\tilde{\xi}\tilde{\xi}}^{-1/2} \tilde{\xi}$, up to quartic order we find 
\begin{align}   
    \label{rnctrans2a}
    \chi  &\; = \; \sqrt{\frac{2}{3 \alpha}}v \tilde{\chi} + \frac{v}{3 \alpha} (\tilde{\chi}^2 - \tilde{\xi}^2 ) + \sqrt{\frac{2}{3}} \frac{v}{9 \alpha^{3/2}} (\tilde{\chi}^3 - 5 \tilde{\chi} \tilde{\xi}^2)+ \frac{v}{54 \alpha^2} (\tilde{\chi}^4 - 18\tilde{\chi}^2 \tilde{\xi}^2 +5\tilde{\xi}^4)~, \\
    \xi & \; = \; \sqrt{\frac{2}{3 \alpha}}v \tilde{\xi} + \frac{2v}{3\alpha} \tilde{\chi} \tilde{\xi} + \sqrt{\frac{2}{3}} \frac{2v}{9 \alpha^{3/2}} (2\tilde{\chi}^2 \tilde{\xi} - \tilde{\xi}^3) + \frac{4v}{27 \alpha^2} (\tilde{\chi}^3 \tilde{\xi} - 2\tilde{\chi} \tilde{\xi}^3) \, . 
    \label{rnctrans2b}
\end{align}
These transformations ensure that the metric is diagonal and canonical. If we use these transformations in the right-hand side of the Lagrangian (\ref{eq:TLag1}), we recover Eq.~(\ref{eq:RNLag2field}).

Next, we show that one can also transform the left-hand side of the Lagrangian~(\ref{eq:TLag1}) to K\"ahler normal coordinates. In this case, the K\"ahler metric is
\begin{equation}
    K_{I\bar{J}} \; = K_{T\bar{T}}=\; \frac{3\alpha}{(T+\bar{T})^2} \, ,
\end{equation}
which leads to the following complex vielbeins
\begin{equation}
    \label{comvielb1}
    e(w, \bar{w}) \; = \; \bar{e}(w, \bar{w}) \; = \; \frac{w+\bar{w}}{\sqrt{3 \alpha}} \; = \; \sqrt{\frac{2}{3\alpha}} v \, .
\end{equation}
We note that the complex vielbein also only depends on the VEV $v$. This choice satisfies the constraint~(\ref{comvielbconst}) and ensures that the K\"ahler metric evaluated at the background is flat.

Assuming the complex field decomposition $T(x) = w+t(x)$, where $w$ is a complex VEV and $t(x)$ is a complex field fluctuation, we can compute the K\"ahler curvature tensor using Eq.~(\ref{kahcurvtens}). We find
\begin{equation}
    R_{t\bar{t}t\bar{t}} \; = \; \frac{6\alpha}{(w+\bar{w})^4} \; = \; \frac{3\alpha}{2v^4} \, .
\end{equation}
The complex vielbein~(\ref{comvielb1}) together with the expression $R_{\alpha \bar{\beta} \gamma \bar{\delta}} = e_{\alpha}{}^{I} \bar{e}_{\bar{\beta}}{}^{\bar{J}} e_{\gamma}{}^K \bar{e}_{\bar{\delta}}{}^{\bar{L}} R_{I\bar{J}K\bar{L}}$, can be used to obtain the K\"ahler curvature tensor
\begin{equation}
    R_{\hat{t}\hat{\bar{t}} \hat{t} \hat{\bar{t}}} \; = \; \frac{2}{3\alpha} \, ,
\end{equation}
where the hatted coordinates denote the K\"ahler normal coordinate basis.

Using the general two-derivative interaction Lagrangian for complex scalar fields expanded in the mass eigenbasis~(\ref{eq:KNClagrangian}), we find that the kinetic terms of the left-hand side of the Lagrangian~(\ref{eq:TLag1}) in the K\"ahler normal coordinates becomes 
\begin{equation}
    \mathcal{K}_{\rm KNC} \; = \;  \partial_\mu \hat{t}\, \partial^{\mu} \hat{\bar{t}}\, +\frac{2}{3\alpha}\hat{t}\hat{\bar{t}}\, \partial_\mu \hat{t}\, \partial^{\mu} \hat{\bar{t}} \, ,
    \label{eq:KNLag2field}
\end{equation}
where the Lagrangian is expanded up to quartic order.

Analogously, instead of obtaining the Lagrangian in K\"ahler normal coordinates with the help of complex vielbeins, we can use a holomorphic transformation~(\ref{eq:transkah1}) together with a field rescaling $\hat{t} \rightarrow \hat{g}_{\hat{t} \hat{\bar{t}}}^{-1/2}\hat{t}$, to obtain the field transformation up to quartic order
\begin{equation}
    \label{holtrans1}
    t \; = \; \sqrt{\frac{2}{3 \alpha}}v \hat{t} + \frac{\sqrt{2}v}{3 \alpha} \hat{t}^2 + \sqrt{\frac{2}{3}} \frac{v}{3 \alpha^{3/2}} \hat{t}^3 + \frac{\sqrt{2}v}{9 \alpha^2} \hat{t}^4 +\dots \; = \; \sqrt{2}v\sum_{n= 1}^{\infty} \left(\frac{\hat{t}}{\sqrt{3\alpha}} \right)^n  \, .
\end{equation}
This field transformation makes the K\"ahler metric diagonal and canonical, and if one applies this transformation on the left-hand side of the Lagrangian~(\ref{eq:TLag1}), one recovers Eq.~(\ref{eq:KNLag2field}).
The amplitude for the complex field has only one independent nonvanishing component, up to crossing/permutation symmetry. It contains two fields and two complex conjugates and is given on-shell by
\begin{equation}
A^{t \, t \, \bar t \, \bar t}_{4, kin} \; = \; \frac{2}{3\alpha}s_{12} \ . \label{eq:complex3}    
\end{equation}
To compute the four-point scattering amplitudes for the real components, we expand the complex field fluctuation $\hat{t} = \frac{1}{\sqrt{2}}\left(\hat{\chi} + i \hat{\xi} \right)$, and the kinetic terms~(\ref{eq:KNLag2field}) in KNC become
\begin{equation}     
    \label{eq:knckinterms2}
    \mathcal{K}_{\rm KNC} \; = \; \frac{1}{2} (\partial_{\mu} \hat{\chi})^2 + \frac{1}{2} (\partial_{\mu} \hat{\xi})^2 + \frac{1}{6\alpha} \left( \hat{\chi}^2 + \hat{\xi}^2\right) \left( (\partial_{\mu} \hat{\chi})^2 + (\partial_{\mu} \hat{\xi})^2 \right) \, .
\end{equation}

We find that the four-point scattering amplitudes in Riemann normal coordinates, given by the kinetic terms~(\ref{eq:RNLag2field}), and in K\"ahler normal coordinates, given by the kinetic terms~(\ref{eq:KNLag2field}), are identical:
\begin{eqnarray}
    A^{\chi \chi \xi \xi}_{4, kin} &\; = \; A^{\xi \xi \chi \chi}_{4, kin} &\; = \; -\frac{2}{3\alpha}s_{12} \, , ~\nonumber\\
    A^{\chi \xi \chi \xi}_{4, kin} &\; = \; A^{\xi \chi \xi \chi}_{4, kin} &\; = \; -\frac{2}{3\alpha}s_{13} \, , ~\nonumber \\
    A^{\chi \xi \xi \chi}_{4, kin} &\; = \; A^{\xi \chi \chi \xi}_{4, kin} &\; = \; -\frac{2}{3\alpha}s_{14} \, ,~\nonumber \\
     A^{\chi \chi \chi \chi}_{4, kin} &\; = \; A^{\xi \xi \xi \xi}_{4, kin} &\; = \; 0 \, . 
\end{eqnarray}
Recall that we are working in Planck units, so the expressions for these amplitudes must be divided by $M_P^2$ so that the amplitudes are dimensionless. Note that we do not include the hats or tildes since the scattering amplitudes coincide in the different bases. Of course, this happens because the potential vanishes, all fields are massless and we have a degenerate vacuum and, as expected, the amplitudes are independent of the VEVs.  Indeed, in the next subsection, we show that the amplitudes only coincide when they are evaluated at an extremum with no tadpoles.    

One can also readily find the transformations between the Riemann and K\"ahler normal coordinates. The transformations between $(\tilde{\chi}, \tilde{\xi}) \leftrightarrow (\hat{\chi}, \hat{\xi})$ up to cubic terms are
\begin{equation}
    \tilde{\chi} \; = \; \hat{\chi} + \frac{1}{18 \alpha} (\hat{\chi} \hat{\xi}^2+ \hat{\chi}^3), \qquad \tilde{\xi} \; = \ \hat{\xi} + \frac{1}{18 \alpha}(\hat{\xi} \hat{\chi}^2 + \hat{\xi}^3) \, ,
\end{equation}
and the inverse transformations
\begin{equation}
   \hat{\chi} \; = \; \tilde{\chi} - \frac{1}{18 \alpha} (\tilde{\chi} \tilde{\xi}^2+ \tilde{\chi}^3), \qquad \hat{\xi} \; = \ \tilde{\xi} - \frac{1}{18 \alpha}(\tilde{\xi} \tilde{\chi}^2 + \tilde{\xi}^3) \, .
\end{equation}

\subsection{General Two-Derivative Interactions}

\subsubsection{Quadratic Potential}
In this subsection, we show how to compute the full four-point interactions when the Lagrangian has both the kinetic and potential terms, $\mathcal{L} = \mathcal{K} - V$. First, we consider a simple quadratic potential of the form $V = m^2 \bar{T} T$. Note that this potential does not arise from a superpotential in no-scale supergravity. 
Instead, the quadratic potential is a simple toy example used to just illustrate the amplitude calculation. In the next subsection we will consider a quadratic superpotential, which leads to a more complicated potential. 

We expand the scalar potential in  terms of real scalar fields using the complex field expansion $T = \frac{1}{\sqrt{2}} \left(X + i Y \right) $,
\begin{equation}
    \label{effpot1}
    V(X, Y) \; = \; \frac{1}{2}m^2 \left(X^2 + Y^2\right) \, .
\end{equation}
Next, decomposing the fields as $X(x) = v + \chi(x)$ and $Y(x) = u + \xi(x)$, and using the field transformation to Riemann normal coordinates from Eqs.~(\ref{rnctrans2a}) and (\ref{rnctrans2b}), we obtain to quartic order
\begin{equation}
     V_{\rm{ RNC}}(\tilde{\chi}, \tilde{\xi}) = m^2v^2\left(\frac{1}{2}+ \sqrt{\frac{2}{3\alpha}}\tilde{\chi} +\frac{2}{3\alpha}{\tilde \chi}^2+\sqrt{\frac{2}{3}}\frac{2}{9\alpha^{3/2}}(2{\tilde \chi}^3-{\tilde \chi}{\tilde \xi}^2) +\frac{4}{27\alpha^2}({\tilde \chi}^4-2{\tilde \chi}^2{\tilde \xi}^2) \right)\,,
     \label{vrncrunaway}
\end{equation}
where, for simplicity, we have set the $Y$ field VEV $u = 0$. However, these computations are completely general and the same steps can be followed for an arbitrary value of $u$. In this case, the masses are $m_{\tilde{\chi}}^2 = 4m^2 v^2/3\alpha$ and $m_{\tilde{\xi}}^2 = 0$. Note that when $v \neq 0$, the potential has a tadpole contribution as seen by the linear term in ${\tilde \chi}$.

One can also expand the complex scalar potential $V = m^2 \bar{T}T$ in terms of its field fluctuation and the VEV, $T(x) = w + t(x)$, and using a holomorphic transformation~(\ref{holtrans1}), we can express the scalar potential in terms of K\"ahler normal coordinates:
\begin{align}
    &V_{\rm{KNC}} (\hat{t}, \hat{\bar{t}}) \; = \;~\nonumber\\ 
    &\frac{m^2 v^2}{2} \left(1 + \frac{2 \hat{t}}{\sqrt{3 \alpha}} + \frac{2\hat{t}^2}{3 \alpha} + 
    \frac{2 \hat{t}^3}{3\sqrt{3} \alpha^{3/2}} + \frac{2 \hat{t}^4}{9 \alpha^2} \right)\left(1 + \frac{2 \hat{\bar{t}}}{\sqrt{3 \alpha}} + \frac{2\hat{\bar{t}}^2}{3 \alpha} + 
    \frac{2 \hat{\bar{t}}^3}{3\sqrt{3} \alpha^{3/2}} + \frac{2 \hat{\bar{t}}^4}{9 \alpha^2} \right)\, .
\end{align}
Alternatively, the scalar potential can be expressed in terms of real scalar field fluctuations and using $\hat{t} = \frac{1}{\sqrt{2}}(\hat{\chi} + i \hat{\xi})$, we obtain
\begin{equation}
    V_{\rm{KNC}} (\hat{\chi}, \hat{\xi}) = m^2 v^2\left(\frac{1}{2}+ \sqrt{\frac{2}{3\alpha}}\hat{\chi} + \frac{2}{3 \alpha} \hat{\chi}^2  + \frac{1}{\sqrt{6}\alpha^{3/2}} (\hat{\chi}^3 -\hat{\chi} \hat{\xi}^2) + \frac{2}{9 \alpha^2} (\hat{\chi}^4 -\hat{\chi}^2 \hat{\xi}^2) \right) \, ,
\end{equation}
where the potential is expanded up to quartic terms. The same expression can be found from Eq.~(\ref{Vknctaylor}) by taking covariant derivatives of the potential and applying the vielbeins. Also note that the masses are $m_{\hat{\chi}}^2 = 4m^2 v^2/3\alpha$ and $m_{\hat{\xi}}^2 = 0$ which is the same as in the RNC case.

We can now calculate the four-point interactions in Riemann normal coordinates, using the Lagrangian $\mathcal{L}_{\rm{RNC}} = \mathcal{K}_{\rm{RNC}} - V_{\rm{RNC}}$, and in K\"ahler normal coordinates, using the Lagrangian $\mathcal{L}_{\rm{KNC}} = \mathcal{K}_{\rm{KNC}} - V_{\rm{KNC}}$. This computation can be performed in two ways.  We can use the perturbative approach using the expressions given in this section. Or analogously, one can use the compact expressions and the vielbein computation to obtain the result.
When $v = 0$, the results are trivial and all amplitudes vanish. When $v \neq 0$, we are in fact attempting to compute the scattering amplitudes when there is no minimum, as it is not possible to eliminate the linear term in Eq.~(\ref{vrncrunaway}). We see that the two approaches (RNC vs KNC) lead to different results, indicating a lack of invariance with respect to field redefinitions. This will be the case whenever the amplitudes are computed away from a minimum.  We summarize the resulting scattering amplitudes in Table~\ref{tab:Pot4pt}. We only include four types of distinct amplitudes, since $A_{4}^{\chi \chi\xi \xi} = A_{4}^{\xi \xi \chi \chi}$ and $A_{4}^{\chi \xi \chi \xi} = A_{4}^{\xi \chi \xi \chi} = A_{4}^{\chi \xi \xi \chi} = A_{4}^{\xi \chi \chi \xi}$. Unlike the case with no potential, the values in Table~\ref{tab:Pot4pt} correspond to the low energy limit of the amplitudes. While we may not expect the individual contributions from kinetic and potential terms to agree in different coordinate systems, the total amplitude should agree in a well defined theory.

\begin{table}[t!]
    \centering
    \begin{tabular}{c|c|c||c||c|c||c||}
    \cline{2-7}
    & \multicolumn{3}{|c||}{\bf{RNC}}&
    \multicolumn{3}{|c||}{\bf{KNC}}\\
    \cline{2-7}
    \hhline{~======}
      & $\mathcal{K}$ & $V$ & Total &  $\mathcal{K}$ & $V$ & Total\\
     \hline
      \multicolumn{1}{|c|}{$\chi \chi \rightarrow \chi \chi$} & $0$ & $\frac{64}{27}$ & $\frac{64}{27}$ & $\frac{16}{9}$ & $\frac{13}{6}$ & $\frac{71}{18}$ \\
         \hline
       \multicolumn{1}{|c|}  {$\xi \xi \rightarrow \xi \xi$} & $0$ & $\frac{8}{27}$ & $\frac{8}{27}$ & $0$ & $\frac{1}{6}$ & $\frac{1}{6}$ \\
      \hline
  \multicolumn{1}{|c|}{$\chi \chi \rightarrow \xi \xi$} & $-\frac{80}{27}$ & $\frac{128}{81}$ & $-\frac{112}{81}$ & $-\frac{8}{3}$ & $\frac{7}{6}$ & $-\frac{3}{2}$ \\
           \hline 
           \multicolumn{1}{|c|}{$\chi \xi \rightarrow \chi \xi$} & $\frac{16}{27}$ & $\frac{32}{81}$ & $\frac{80}{81}$ & $\frac{8}{9}$ & $\frac{5}{18}$ & $\frac{7}{6}$ \\\hline
    \end{tabular} 
    \caption{Four-point scattering amplitudes in the nonrelativistic limit in units of $m^2 v^2/M_P^4\alpha^2$ for the case of a simple mass term. For $v\neq 0$, the amplitudes do not match between RNC and KNC as there are no extremal points in the (canonical) potential.}
    \label{tab:Pot4pt}
\end{table}

\subsubsection{Quadratic Superpotential}
\label{sec:quadrsup1}
\begin{figure}[t!]
    \centering
    \includegraphics[width=0.8\columnwidth]{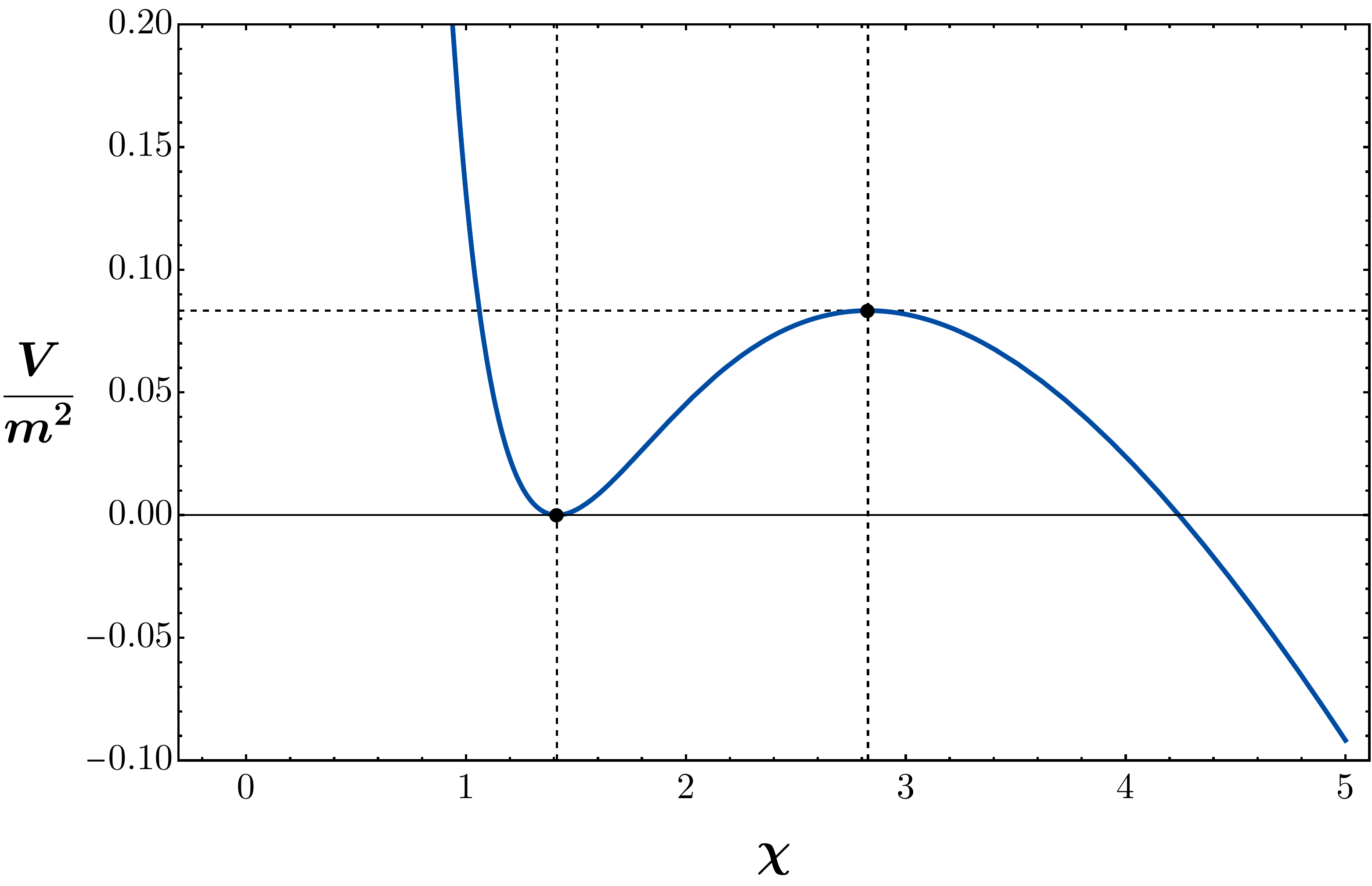}
    \caption{The potential (\ref{eq:V2pot}) showing the extrema $(v,V/m^2)$ at $(\sqrt{2},0)$ and $(2\sqrt{2}, \frac{1}{12})$. }
    \label{fig:V2}
\end{figure}

We next study a model that can arise naturally in a no-scale supergravity framework with the following quadratic superpotential
\begin{equation}
    W(T) = m (T - 1)^2 \, .
    \label{eq:spT}
\end{equation}
Assuming the same no-scale K\"ahler potential and metric~(\ref{kah1}), the superpotential (\ref{eq:spT}) leads to the following scalar potential:
\begin{eqnarray}
    V(T,\bar T)&=&\frac{m^2(T-1)(\bar T-1)}{(T+\bar T)^{3\alpha}}\left[\left(\frac{4}{3\alpha}-2\right)(T+\bar T)^2+4 (T+\bar T)+3(\alpha-1) (T-1)(\bar T-1) \right]~\nonumber\\
    &=& -\frac{2}{3}m^2\frac{(T-1)(\bar T-1)}{(T+\bar T)^2}(T+\bar T-6)~, \qquad\qquad (\alpha=1) \, .
    \label{eq:V2pot}
\end{eqnarray}
In terms of real fields $T = \frac{1}{\sqrt{2}}(\chi + i \xi)$, this potential becomes
\begin{equation}
    V(\chi, \xi) \; = \; \frac{m^2}{6 \chi ^2} \left(6 \left(2+\xi ^2\right)-\sqrt{2}\chi \left(14+\xi ^2\right) +10 \chi ^2 -\sqrt{2} \chi ^3\right) \, .
     \label{eq:V2potreal}
\end{equation}
This potential is plotted  in Figure~\ref{fig:V2} for $\alpha = 1$ and $\xi = 0$. There are two extrema, one at $\langle \chi \rangle = v = \sqrt{2}$ (local minimum) and $\langle \chi \rangle = v = 2\sqrt{2}$ (local maximum).

Using the transformations \eqref{rnctrans2a}, \eqref{rnctrans2b} and assuming $\alpha=1$, one obtains the potential in Riemann normal coordinates
\begin{eqnarray}
     &&V_{\rm RNC}(\tilde{\chi}, \tilde{\xi})=\frac{m^2}{6 v^2} \left(12-v \left(v \left(\sqrt{2} v-10\right)+14 \sqrt{2}\right)\right)-\frac{m^2}{3 \sqrt{3} v^2} \left(v^3-14 v+12 \sqrt{2}\right) \tilde{\chi} \nonumber\\
     &&-\frac{m^2}{18 v^2} \left(\left(\sqrt{2} v \left(v^2+14\right)-48\right) \tilde{\chi}^2 - \left(24-v \left(v \left(\sqrt{2} v-12\right)+14 \sqrt{2}\right)\right)\tilde{\xi}^2 \right)~\nonumber \\
     &&-\frac{m^2}{27 \sqrt{3} v^2} \left(v^3-14 v+48 \sqrt{2}\right) \tilde{\chi}  \left(\tilde{\xi}^2+\tilde{\chi}^2\right)~\nonumber \\
     &&-\frac{m^2}{324 v^2} \left(\left(\sqrt{2} v \left(v^2+14\right)-192\right) \tilde{\chi} ^2+\tilde{\xi}^2 \left(v \left(v \left(\sqrt{2} v-48\right)+14 \sqrt{2}\right)-96\right)\right) \left(\tilde{\xi} ^2+\tilde{\chi} ^2\right)~\nonumber \\
     &&= \frac{2m^2}{81} (\tilde{\chi}^2 + \tilde{\xi}^2) \left(18 - 9\sqrt{6} \tilde{\chi} + 10 \tilde{\chi}^2 + 10 \tilde{\xi}^2 \right) \, ,
    \label{eq:potalpha1}
\end{eqnarray}
where in the last line we have taken $v = \sqrt{2}$, corresponding to the potential minimum. 

The potential in K\"ahler normal coordinates can also be obtained. Using the KNC transformation~(\ref{holtrans1}) together with the potential~(\ref{eq:V2potreal}), we find 
\begin{eqnarray}
     &&V_{\rm KNC}(\hat{\chi}, \hat{\xi})=\frac{m^2}{6 v^2} \left(12-v \left(v \left(\sqrt{2} v-10\right)+14 \sqrt{2}\right)\right)-\frac{m^2}{3 \sqrt{3} v^2} \left(v^3-14 v+12 \sqrt{2}\right) \hat{\chi} \nonumber\\
     &&-\frac{m^2}{18 v^2} \left(\left(\sqrt{2} v \left(v^2+14\right)-48\right) \hat{\chi}^2 - \left(24-v \left(v \left(\sqrt{2} v-12\right)+14 \sqrt{2}\right)\right)\hat{\xi}^2 \right)~\nonumber \\
     &&-\frac{m^2}{18 \sqrt{3} v^2} \left(v^3-14 v+36 \sqrt{2}\right) \hat{\chi}  \left(\hat{\xi} ^2+\hat{\chi}^2\right)~\nonumber\\
     &&-\frac{m^2}{108 v^2} \left(\left(\sqrt{2} v \left(v^2+14\right)-96\right) \hat{\chi}^2+\hat{\xi}^2 \left(v \left(v \left(\sqrt{2} v-24\right)+14 \sqrt{2}\right)-48\right)\right)\left(\hat{\xi}^2+\hat{\chi}^2\right) ~\nonumber \\
     &&=\frac{2m^2}{27}(\hat{\chi}^2 + \hat{\xi}^2)(6 - 3\sqrt{6} \hat{\chi} + 4 \hat{\chi}^2 + 4 \hat{\xi}^2) \, ,
    \label{eq:potKNC}
\end{eqnarray}
where, as before, in the last line we have set $v = \sqrt{2}$.

Comparing the expressions \eqref{eq:potKNC} and \eqref{eq:potalpha1} at the local minimum ($v=\sqrt{2}$), we see that the KNC potential differs from the RNC potential only in the quartic interactions. In both cases, the masses at the local minimum are degenerate and given by $m_{\tilde{\chi}}^2 = m_{\hat{\chi}}^2 = \frac{8m^2}{9}$ and $m_{\tilde{\xi}}^2 = m_{\hat{\xi}}^2 = \frac{8m^2}{9}$. Similarly, at the local maximum the masses are again degenerate with $m_{\tilde{\chi}}^2 = m_{\hat{\chi}}^2 = -\frac{5m^2}{9}$ and $m_{\tilde{\xi}}^2 = m_{\hat{\xi}}^2 = \frac{4m^2}{9}$.

\begin{figure}[ht!]
    \centering
    \includegraphics[width=0.8\columnwidth]{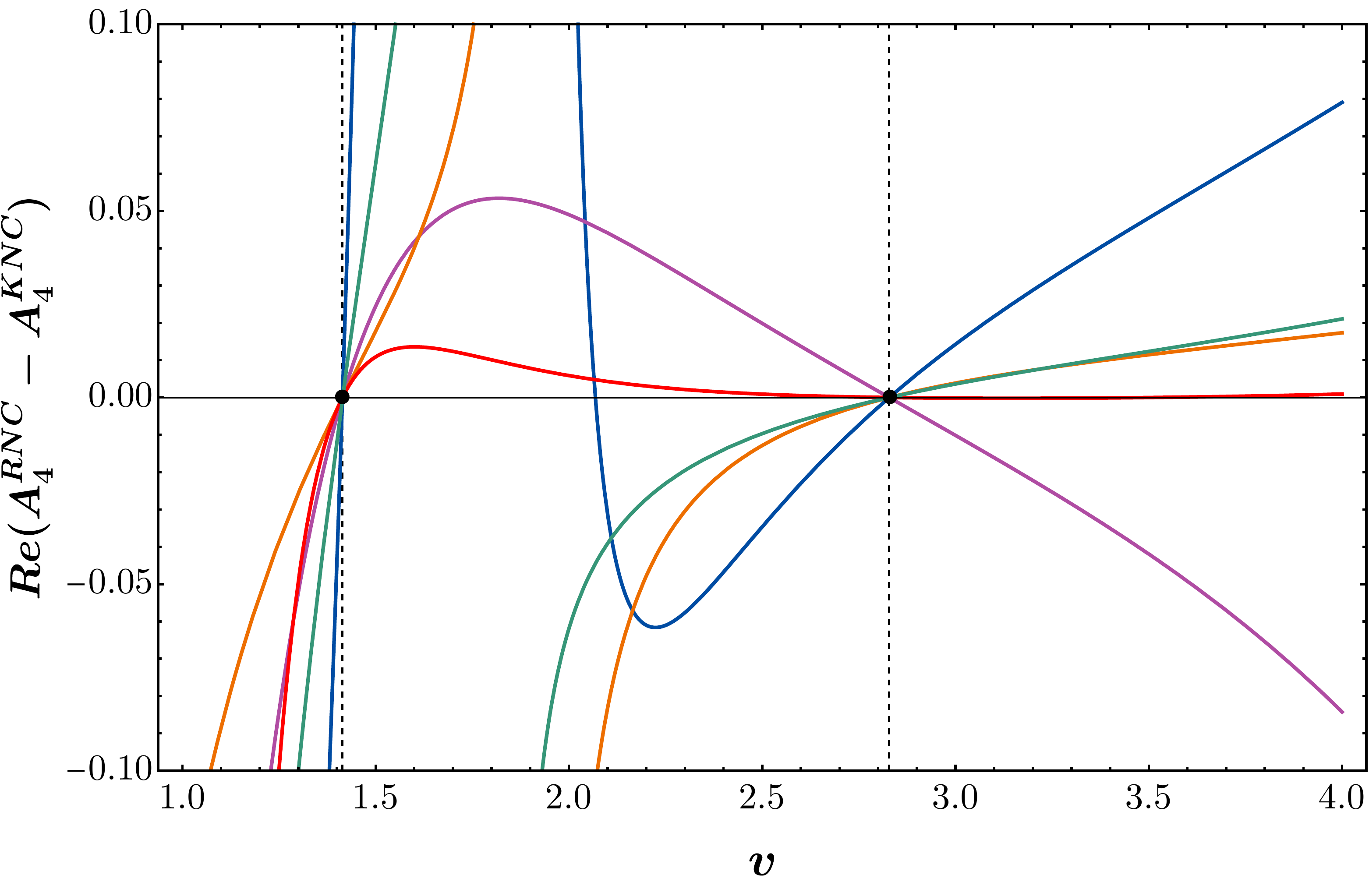}
        \caption{The difference between the real values of RNC and KNC four-point amplitudes for the quadratic superpotential \eqref{eq:spT} as a function of $v$ for each channel (blue: $\chi \chi \rightarrow \chi \chi$, purple: $\xi \xi \rightarrow \xi \xi$, orange: $\chi \chi \rightarrow \xi \xi$, red: $\xi \xi \rightarrow \chi \chi$, green: $\chi \xi \rightarrow \chi \xi$). We assume $\alpha = 1$ and units of $m^2/M_P^2$. The amplitudes only agree at $v = \sqrt{2}, \, 2\sqrt{2}$ and $-3\sqrt{2}$ (not shown), corresponding to extrema of the potential.}
    \label{fig:ampdiffs}
\end{figure}

The potentials \eqref{eq:potalpha1} and \eqref{eq:potKNC}, together with the kinetic terms in RNC and KNC given by (\ref{eq:RNLag2field}) and (\ref{eq:knckinterms2}), respectively, can now be used to compute the four-point scattering amplitudes. The amplitudes are calculated by ignoring the linear term in the fluctuation which are of course absent at the extrema. For each channel, the amplitudes only agree at $v=\sqrt{2}$ and $2\sqrt{2}$, corresponding to the extrema of the potential\footnote{There is an extremum at $v=-3\sqrt{2}$ for which the amplitudes also match but we only consider field values $v>0$ that are consistent with the K\"ahler potential \eqref{kah1}.}, or equivalently, where the coefficient of the linear term vanishes. This behaviour is depicted in Figure~\ref{fig:ampdiffs} and the values of the amplitudes for $v=\sqrt{2}$ (top row)  and $2\sqrt{2}$ (bottom row) are given in Table~\ref{tab:Potminscattering}. To obtain numerical values of the scattering amplitudes, we choose $s_{12} = (m_1 + m_2)^2$ and $s_{13} = s_{14}$, where $s_{12} + s_{13} + s_{14} = \sum_{i=1}^4 m_i^2$ is always satisfied. However, since for some values of $v$ the amplitudes become complex, only the real part of the amplitude is plotted. Importantly, Fig.~\ref{fig:ampdiffs} illustrates that the scattering amplitudes evaluated using different coordinates (i.e. RNC and KNC) only coincide at the potential extrema \footnote{We further caution that our calculations of the amplitudes at the maximum still assume a Minkowski background. Though this is clearly not correct and a full computation in a de Sitter background is far beyond the scope of this paper, we expect the agreement between KNC and RNC to be sustained.}.

\begin{table}[ht!]
    \centering
    \begin{tabular}{c|c|c||c||c|c||c||}
    \cline{2-7}
    & \multicolumn{3}{|c||}{\bf{RNC}}&
    \multicolumn{3}{|c||}{\bf{KNC}}\\
    \cline{2-7}
    \hhline{~======}
      & $\mathcal{K}$ & $V$ & Total &  $\mathcal{K}$ & $V$ & Total\\
      \hline
      \multicolumn{1}{|c|}{$\chi \chi \rightarrow \chi \chi$} & $0$ & $\frac{380}{27}$ & $\frac{380}{27}$ & $\frac{32}{27}$ & $\frac{116}{9}$ & $\frac{380}{27}$ \\
      \multicolumn{1}{|c|} {}  & $0$ & $-\frac{80}{27}$ & $-\frac{80}{27}$ & $-\frac{20}{27}$ & $-\frac{20}{9}$ & $-\frac{80}{27}$ \\
         \hline
       \multicolumn{1}{|c|}  {$\xi \xi \rightarrow \xi \xi$} & $0$ & $-\frac{100}{27}$ & $-\frac{100}{27}$ & $\frac{32}{27}$ & $-\frac{44}{9}$ & $-\frac{100}{27}$ \\
      \multicolumn{1}{|c|}{}   & $0$ & $-\frac{3712}{945}$ & $-\frac{3712}{945}$ & $\frac{16}{27}$ & $-\frac{1424}{315}$ & $-\frac{3712}{945}$ \\
      \hline
\multicolumn{1}{|c|}{$\chi \chi \rightarrow \xi \xi$} & $-\frac{128}{81}$ & $-\frac{52}{81}$ & $-\frac{20}{9}$ & $-\frac{32}{27}$ & $-\frac{28}{27}$ & $-\frac{20}{9}$ \\
  \multicolumn{1}{|c|}{}
           & $\frac{116}{81}$ & $-\frac{364}{405}$ & $\frac{8}{15}$ & $\frac{38}{27}$ & $-\frac{118}{135}$ & $\frac{8}{15}$ \\
           \hline 
           \multicolumn{1}{|c|}{$\xi \xi \rightarrow \chi \chi$} & $-\frac{128}{81}$ & $-\frac{52}{81}$ & $-\frac{20}{9}$ & $-\frac{32}{27}$ & $-\frac{28}{27}$ & $-\frac{20}{9}$ \\
    \multicolumn{1}{|c|}{}
           & $-\frac{100}{81}$ & $-\frac{5588}{7371}$ & $-\frac{544}{273}$ & $-\frac{34}{27}$ & $-\frac{1802}{2457}$ & $-\frac{544}{273}$ \\
           \hline 
           \multicolumn{1}{|c|}{$\chi \xi \rightarrow \chi \xi$} & $\frac{64}{81}$ & $\frac{236}{81}$ & $\frac{100}{27}$ & $\frac{32}{27}$ & $\frac{68}{27}$ & $\frac{100}{27}$ \\
           \multicolumn{1}{|c|}{}
           & $-\frac{1}{81}$ & $-\frac{11512}{13041}$ & $-\frac{1297}{1449}$ & $-\frac{1}{27}$ & $-\frac{3730}{4347}$ & $-\frac{1297}{1449}$ \\\hline
    \end{tabular}
    \caption{Four-point scattering amplitudes evaluated using RNC and KNC in the limit $s_{12} = (m_1 + m_2)^2$ and $s_{13} = s_{14}$ in units of $m^2/M_P^2$ with $\alpha = 1$ for the case of a potential arising from the superpotential \eqref{eq:spT}. For each channel, upper (lower) entries are for $v=\sqrt{2}~(2 \sqrt{2})$ corresponding to the local minimum (maximum) of the potential. Note that for the $\chi\xi$ channel at the maximum $(v = 2\sqrt{2})$, the amplitude matches as an imaginary number, but only the real value is quoted in the Table.}
    \label{tab:Potminscattering}
\end{table}

For other (non-extremal) values of $v$, the linear term will be nonzero and needs to be included in the four-point scattering. We know of no way to include all of the tadpole contributions generally. It was however shown in \cite{Dudas:2004nd} for simple examples that physical quantities computed away from extrema, obtained by combining the naive result with a full resummation of tree-level tadpole diagrams, gives the same results as if computed at an extremum field value. In agreement with this, in certain cases where the tadpole contribution can be computed perturbatively, their sum leads to the same result for the scattering amplitudes as computed at the minimum \cite{dgov2}. The RNC values in Table~\ref{tab:Potminscattering} can be obtained either by using the Feynman rules following from the Lagrangian $\mathcal{L}_{\rm RNC} = \mathcal{K}_{\rm RNC} - V_{\rm RNC}$ or directly using the expression \eqref{eq:A4amp}. Analogously, the KNC values in Table~\ref{tab:Potminscattering} can be found either by using the Feynman rules following from the Lagrangian $\mathcal{L}_{\rm KNC} = \mathcal{K}_{\rm KNC} - V_{\rm KNC}$ or directly using the expressions \eqref{Amp4kin:KNC} and \eqref{Amp4:KNC}. From the table entries, we see that the total amplitudes are invariant with respect to the choice of coordinates 
(RNC or KNC), although the particular kinetic or potential contribution is not.

\subsection{KKLT}
\begin{figure}[t]
    \centering
    \includegraphics[width=0.8\columnwidth]{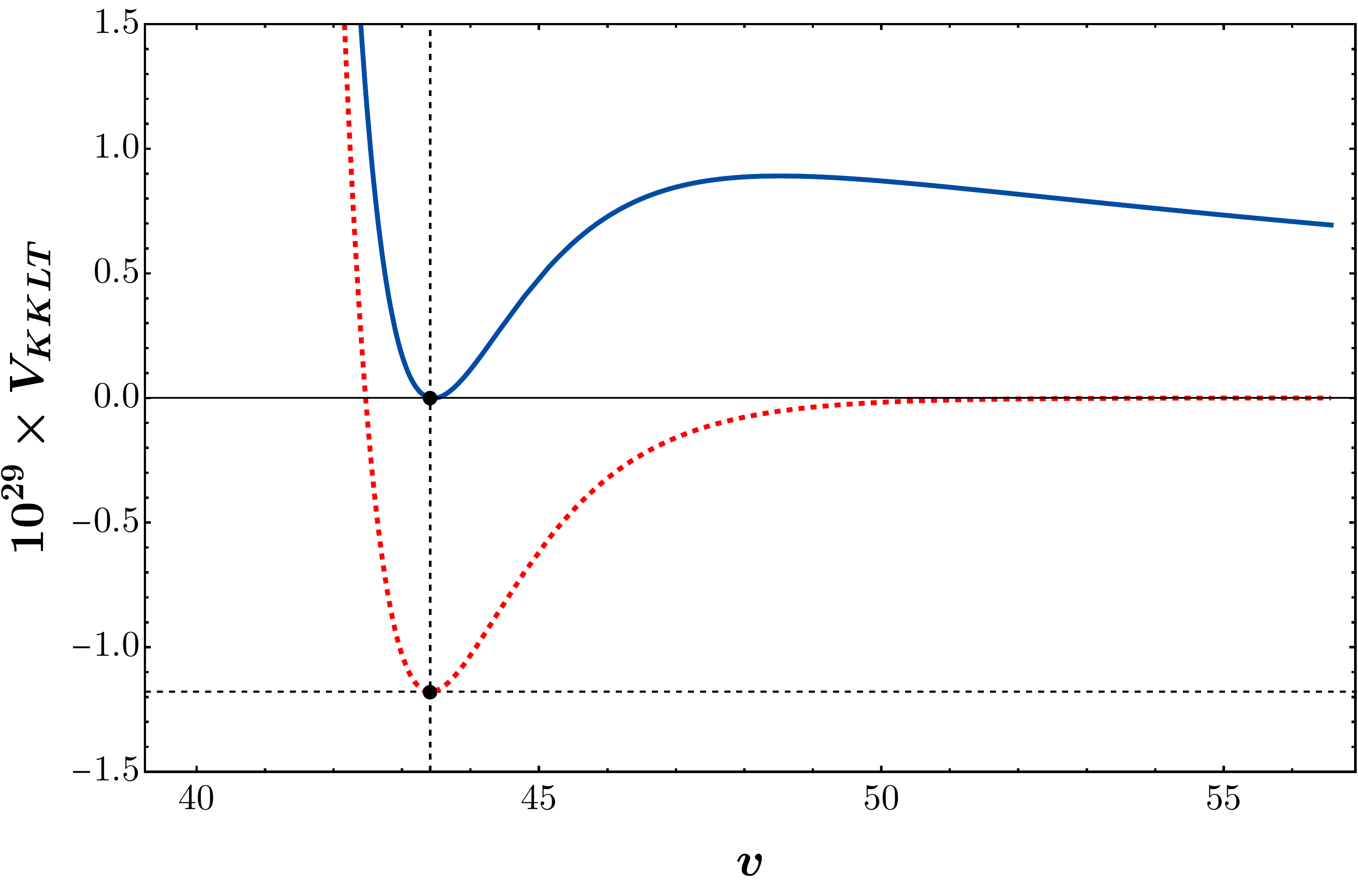}
    \caption{The KKLT potential with no uplift (dashed red) and the uplifted potential (solid blue) corresponding to the parameters $B=1, b=1, W_0=-10^{-12}$, and $\alpha=1$. The KKLT potential has an AdS minimum at $v=43.4127$, while the uplifted potential has a minimum at $v=43.4573$ and a maximum at $v=48.5168$. }
    \label{fig:VKKLT}
\end{figure}

Perhaps a more realistic potential for $T$ comes from stabilizing the modulus with the KKLT superpotential~{\cite{Kachru:2003aw}}
\beq
\label{sup:KKLT}
W = W_0 + B e^{-b T} \, ,
\eeq
where $W_0$ and $b > 0$ are constants. In this model, there is a supersymmetry preserving anti-de Sitter (AdS) minimum which can be found by setting the covariant derivative $D_T$ to zero, namely
\beq
 W_T + W K_T  = 0 \, .
\eeq
For a given constant, $W_0$, this implicitly determines the expectation value of $T = {\bar T}=v/\sqrt{2}$ at the minimum via the equation
\beq
W_0 \; = \; -B e^{-\frac{b v}{\sqrt{2}}} \left(1 + \frac{\sqrt{2}}{3} b v \right) \, .
\eeq
Using the scalar potential expression (\ref{eq:effpotsugra}) with the KKLT superpotential~(\ref{sup:KKLT}), the full scalar potential is given by
\beq
V_{\rm KKLT} = \frac{b B \, e^{-b (T + {\bar T}) }}{3 (T + {\bar T})^2} \left( 3 (e^{b T}+e^{b {\bar T}})W_0 + B (6 + b(T +{\bar T})) \right)\,.
\eeq
At the minimum \cite{Linde:2011ja},  
\beq
V_{\rm AdS} = - \frac{1}{3\sqrt{2} v} b^2 B^2 e^{-\sqrt{2} b v} \, .
\eeq 
This AdS minimum must be uplifted to an approximately zero value, 
which induces supersymmetry breaking. This can be achieved by adding the potential term
\begin{equation}
    \Delta V \approx |V_{\rm AdS}| \left(\frac{\sqrt{2} v^{\rm AdS}}{T + {\bar T}}\right)^2\,,
\end{equation}
where $v^{\rm AdS}$ is the VEV corresponding to the  AdS minimum of $V_{\rm KKLT}$. The uplifted potential, $V_{\rm KKLT} +\Delta V$ still has a minimum (now with $V \approx 0$) but now also has a maximum. The position of the minimum shifts slightly after uplifting with $\Delta v/v = 2/(b v)^2$ 
for $b v \gg 1$. These potentials are depicted in Figure~\ref{fig:VKKLT} for the choices $b = B = 1$ and $W_0 = -10^{-12}$ \cite{Linde:2011ja}.

\begin{figure}[ht!]
    \centering
    \includegraphics[width=0.75\columnwidth]{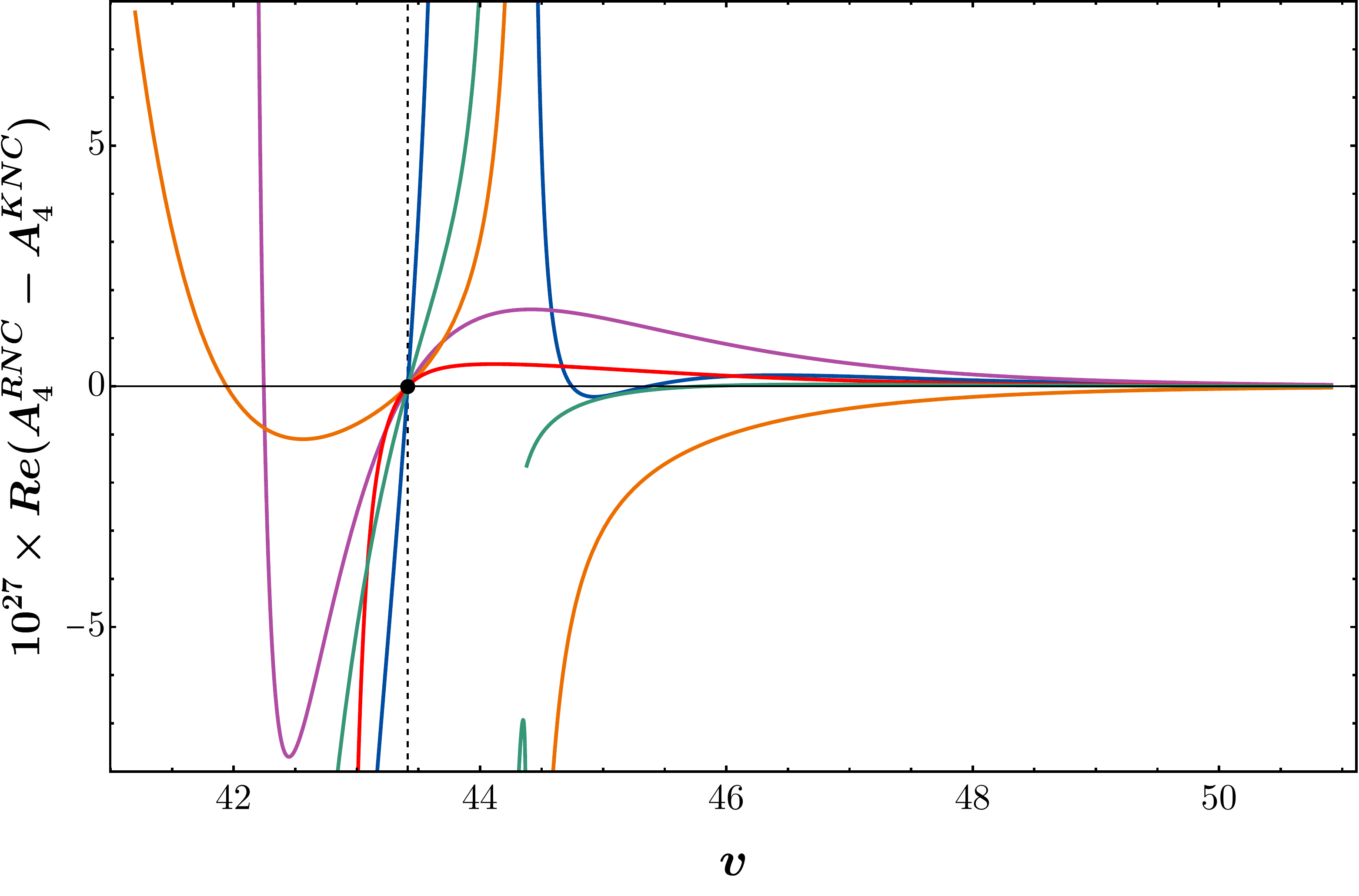}
    \includegraphics[width=0.75\columnwidth]{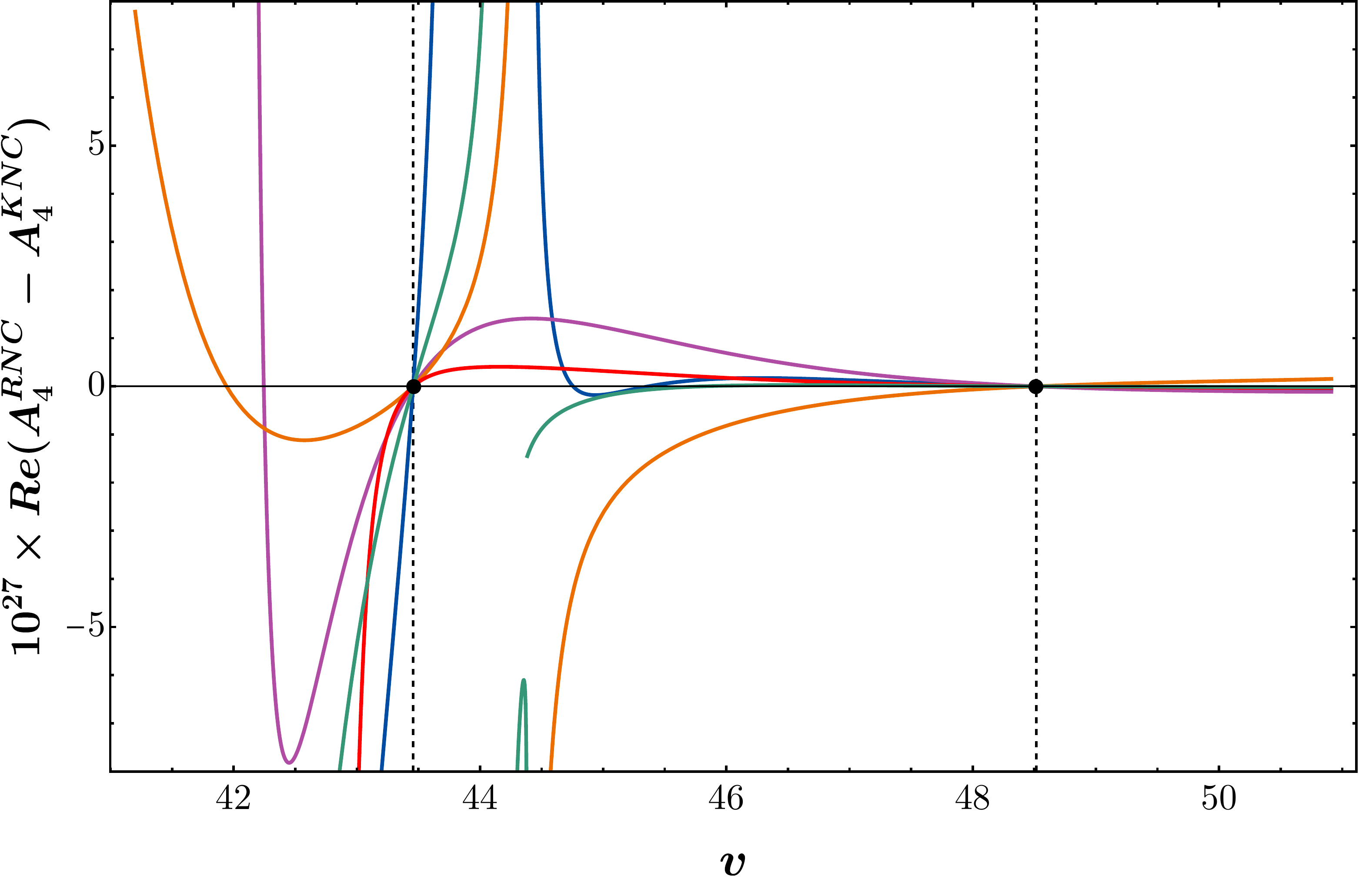}
    \caption{The difference between the RNC and KNC four-point amplitudes for the KKLT potential (top) and the potential with uplift (bottom) as a function of $v$ for each channel (blue: $\chi \chi \rightarrow \chi \chi$, purple: $\xi \xi \rightarrow \xi \xi$, orange: $\chi \chi \rightarrow \xi \xi$, red: $\xi \xi \rightarrow \chi \chi$, green: $\chi \xi \rightarrow \chi \xi$). Assuming $B=1,b=1, W_0=-10^{-12}$ and $\alpha=1$, the amplitudes in the top figure only agree at the AdS minimum $v^{\rm AdS}=43.4127$, while in the bottom figure the amplitudes agree at both the minimum $v_0=43.4573$ and maximum $v=48.5168$ of the potential.     
    }
    \label{fig:KKLTampdiffs}
\end{figure}

The potentials can be converted to RNC and KNC, and we study the four-point scattering amplitudes as a function of $v$. The results are shown in Figure~\ref{fig:KKLTampdiffs} for both $V_{\rm KKLT}$ and the uplifted potential. Note that just as in the previous section, the scattering amplitudes are calculated at the particular values, $s_{12} = (m_1 + m_2)^2$ and $s_{13} = s_{14}$ (assuming $s_{12} + s_{13} + s_{14} = \sum_{i=1}^4 m_i^2$), and only the real part of the amplitude is shown in the figure. For $V_{\rm KKLT}$ (which just has a minimum) the amplitudes for the four channels only agree at the AdS minimum of the KKLT potential where the linear term in the potential vanishes. Away from the minimum the amplitudes do not agree since the linear term has been ignored. Instead for the uplifted potential (which has both a minimum and maximum) the amplitudes not only agree at the minimum but also at maximum of the uplifted potential.

\subsection{No-Scale models including Matter Fields}
\label{su21}

We next turn to the somewhat more realistic case where matter fields $\phi$ are included along with the modulus, $T$.
The K\"ahler potential  in this case is given by Eq.~(\ref{v0}).
As we will see, adding a single matter field greatly increases
the complexity of the calculation whether we are considering
2 complex fields or 4 real fields. The metric is no longer diagonal and the transformations to RNC and KNC are significantly more complicated. Nevertheless, analytic expressions are still manageable as we show below. 
As in the previous case of a single complex field,
we first treat the Lagrangian with the absence of a potential
before adding a  simple cubic interaction for the matter field in the superpotential.

\subsubsection{Two Field Two-Derivative Interactions with No Potential}
Our starting point is the general supergravity Lagrangian~(\ref{lagsugra}), that we combine with the K\"ahler potential~(\ref{v0}) for two complex fields. We expand the kinetic terms of the Lagrangian,
\begin{align}
	\label{lag2fields}
    \mathcal{L} \; = \; & K_{I \bar{J}}(Z, \bar{Z}) \partial_{\mu}Z \partial^{\mu}\bar{Z} \; ~\nonumber\\
    =\; & \frac{\alpha}{\left( T + \bar{T}- \frac{|\phi|^2}{3} \right)^2} \left(3\partial_{\mu} T \partial^{\mu} \bar{T} - \phi \partial_{\mu}T \partial^{\mu} \bar{\phi}  - \bar{\phi} \partial_{\mu} \bar{T} \partial^{\mu} \phi + (T + \bar{T}) \partial_{\mu} \phi \partial^{\mu} \bar{\phi}
    \right)
    \, ,
\end{align}
where $Z = \{T, \phi \}$, and the $SU(2,1)$ K\"ahler metric is given by
\begin{equation}
	\label{kahmetric2fields}
    K_{I \bar{J}} \; = \; \frac{\alpha}{\left( T + \bar{T}- \frac{|\phi|^2}{3} \right)^2}
    \begin{pmatrix}
    3 & -\phi \\
    -\bar{\phi} & T + \bar{T}
    \end{pmatrix} \, .
\end{equation}
Next, we expand the complex scalar fields $T = \frac{1}{\sqrt{2}}(X + i Y)$, $\phi = \frac{1}{\sqrt{2}}(P + i S)$, and the kinetic terms of the Lagrangian become
\begin{align}
    \mathcal{L} \; = \; \frac{36 \alpha}{\left(6\sqrt{2} X- P^2 - S^2 \right)^2} \left(\frac{3}{2} \left((\partial_{\mu} X)^2 + (\partial_{\mu} Y)^2 \right) - \frac{P}{\sqrt{2}}(\partial_{\mu}Y \partial^{\mu}S + \partial_{\mu}P \partial^{\mu}X) \right.~\nonumber\\+ 
    \left. \frac{S}{\sqrt{2}} \left(\partial_{\mu}Y\partial^{\mu}P - \partial_{\mu}S \partial^{\mu}X \right) + \frac{X}{\sqrt{2}} \left((\partial_{\mu} P)^2 + (\partial_{\mu} S)^2 \right) \right)
    \, .
\end{align}
Equivalently, one can use the compact expression~$\mathcal{L} =\frac{1}{2} G_{AB} \partial_{\mu} \Phi^{A} \partial^{\mu} \Phi^B$, given by Eq~(\ref{eqgencomp}), with $\Phi = \{ X, P, Y, S \}$, where the real metric is given by
\begin{equation}
	\label{twofieldmetr}
    G_{AB} \; = \;  \frac{36 \alpha}{\left(6\sqrt{2} X - P^2 - S^2 \right)^2}
    \begin{pmatrix}
    3 & -\frac{P}{\sqrt{2}} & 0 & -\frac{S}{\sqrt{2}} \\ 
    -\frac{P}{\sqrt{2}} & \sqrt{2} X & \frac{S}{\sqrt{2}} & 0 \\
    0 & \frac{S}{\sqrt{2}} & 3 & -\frac{P}{\sqrt{2}} \\ 
    -\frac{S}{\sqrt{2}} & 0 & -\frac{P}{\sqrt{2}} & \sqrt{2} X
    \end{pmatrix} 
    \, .
\end{equation}
To simplify the analysis, we expand the real fields in terms of their field fluctuations and VEVs, $X(x) = v+\chi(x)$, $Y(x) = u+\xi(x)$, $P = p+\rho(x)$, and $S = s+\sigma(x)$, and set the VEVs $u=p = s = 0$. In this case, the metric~(\ref{twofieldmetr}) evaluated at the background VEV takes the simple form
\begin{equation}
    G_{AB}(v) \; = \;
    \begin{pmatrix}
    \frac{3\alpha}{2v^2} & 0 & 0 & 0 \\ 
    0 & \frac{\alpha}{\sqrt{2}v} & 0 & 0 \\
    0 & 0 & \frac{3\alpha}{2v^2} & 0 \\ 
    0 & 0 & 0 & \frac{\alpha}{\sqrt{2}v}
    \end{pmatrix} 
    \, ,
\end{equation}
and the vielbein, that satisfies the constraint $\delta_{ab} = e_a{}^A e_b{}^B G_{AB}$, is
\begin{equation}
    e_a^A(v) \; = \; 
    \begin{pmatrix}
    \sqrt{\frac{2}{3\alpha}} v & 0 & 0 & 0 \\
    0 & \sqrt{\frac{\sqrt{2}v}{\alpha}} & 0 & 0 \\
    0 & 0 & \sqrt{\frac{2}{3\alpha}} v & 0 \\
    0 & 0 & 0 & \sqrt{\frac{\sqrt{2}v}{\alpha}}
    \end{pmatrix} \, .
\end{equation}
Using the expression~(\ref{riemtensor}), we find that the non-zero curvature components in the mass eigenbasis evaluated at the flat background are given by
\begin{eqnarray}
    -R_{\tilde{\chi} \tilde{\rho} \tilde{\chi} \tilde{\rho} } = R_{\tilde{\chi} \tilde{\rho} \tilde{\rho} \tilde{\chi} }= -R_{\tilde{\chi} \tilde{\rho} \tilde{\xi} \tilde{\sigma} } = R_{\tilde{\chi} \tilde{\rho} \tilde{\sigma} \tilde{\xi} } = -R_{\tilde{\chi} \tilde{\sigma} \tilde{\chi} \tilde{\sigma} } = -R_{\tilde{\chi} \tilde{\sigma} \tilde{\rho} \tilde{\xi} } = R_{\tilde{\chi} \tilde{\sigma} \tilde{\xi} \tilde{\rho} } = R_{\tilde{\chi} \tilde{\sigma} \tilde{\sigma} \tilde{\chi} }\nonumber \\
    = R_{\tilde{\rho} \tilde{\chi} \tilde{\chi} \tilde{\rho}} =  -R_{\tilde{\rho} \tilde{\chi} \tilde{\rho} \tilde{\chi} } = R_{\tilde{\rho} \tilde{\chi} \tilde{\xi} \tilde{\sigma} }= -R_{\tilde{\rho} \tilde{\chi} \tilde{\sigma} \tilde{\xi} } = -R_{\tilde{\rho} \tilde{\xi} \tilde{\chi} \tilde{\sigma} }  = -R_{\tilde{\rho} \tilde{\xi} \tilde{\rho} \tilde{\xi} } =  R_{\tilde{\rho} \tilde{\xi} \tilde{\xi} \tilde{\rho} }  = R_{\tilde{\rho} \tilde{\xi} \tilde{\sigma} \tilde{\chi} }  \nonumber \\
    =R_{\tilde{\xi} \tilde{\rho} \tilde{\chi} \tilde{\sigma} }  =  R_{\tilde{\xi} \tilde{\rho} \tilde{\rho} \tilde{\xi} } = -R_{\tilde{\xi} \tilde{\rho} \tilde{\xi} \tilde{\rho} }  = -R_{\tilde{\xi} \tilde{\rho} \tilde{\sigma} \tilde{\chi} } =  -R_{\tilde{\xi} \tilde{\sigma} \tilde{\chi} \tilde{\rho} } =  R_{\tilde{\xi} \tilde{\sigma} \tilde{\rho} \tilde{\chi} }  =  -R_{\tilde{\xi} \tilde{\sigma} \tilde{\xi} \tilde{\sigma} } =  R_{\tilde{\xi} \tilde{\sigma} \tilde{\sigma} \tilde{\xi} }\nonumber\\
   =R_{\tilde{\sigma} \tilde{\chi} \tilde{\chi} \tilde{\sigma} } = R_{\tilde{\sigma} \tilde{\chi} \tilde{\rho} \tilde{\xi} }=-R_{\tilde{\sigma} \tilde{\chi} \tilde{\xi} \tilde{\rho} }=-R_{\tilde{\sigma} \tilde{\chi} \tilde{\sigma} \tilde{\chi} } = R_{\tilde{\sigma} \tilde{\xi} \tilde{\chi} \tilde{\rho} } = -R_{\tilde{\sigma} \tilde{\xi} \tilde{\rho} \tilde{\chi} } = R_{\tilde{\sigma} \tilde{\xi} \tilde{\xi} \tilde{\sigma} } =  -R_{\tilde{\sigma} \tilde{\xi} \tilde{\sigma} \tilde{\xi} } =  \frac{1}{6 \alpha} ,\\
    -R_{\tilde{\chi} \tilde{\xi} \tilde{\chi} \tilde{\xi} }=R_{\tilde{\chi} \tilde{\xi} \tilde{\xi} \tilde{\chi} }=-R_{\tilde{\rho} \tilde{\sigma} \tilde{\rho} \tilde{\sigma} }=R_{\tilde{\rho} \tilde{\sigma} \tilde{\sigma} \tilde{\rho} }=R_{\tilde{\xi} \tilde{\chi} \tilde{\chi} \tilde{\xi} }=-R_{\tilde{\xi} \tilde{\chi} \tilde{\xi} \tilde{\chi} }=R_{\tilde{\sigma} \tilde{\rho} \tilde{\rho} \tilde{\sigma} }=-R_{\tilde{\sigma} \tilde{\rho} \tilde{\sigma} \tilde{\rho} } =  \frac{2}{3\alpha}, \\
    -R_{\tilde{\chi} \tilde{\xi} \tilde{\rho} \tilde{\sigma} }=R_{\tilde{\chi} \tilde{\xi} \tilde{\sigma} \tilde{\rho} }=-R_{\tilde{\rho} \tilde{\sigma} \tilde{\chi} \tilde{\xi} }=R_{\tilde{\rho} \tilde{\sigma} \tilde{\xi} \tilde{\chi} }=R_{\tilde{\xi} \tilde{\chi} \tilde{\rho} \tilde{\sigma} }=-R_{\tilde{\xi} \tilde{\chi} \tilde{\sigma} \tilde{\rho} }=R_{\tilde{\sigma} \tilde{\rho} \tilde{\chi} \tilde{\xi} }=-R_{\tilde{\sigma} \tilde{\rho} \tilde{\xi} \tilde{\chi} } = \frac{1}{3\alpha} \, ,
\end{eqnarray}
where, as before, tilde denotes the Riemann normal coordinate basis.

Using the general two-derivative interaction Lagrangian expansion in the mass eigenbasis~(\ref{lagrnc}), the kinetic terms of the Lagrangian~(\ref{lag2fields}) in Riemann normal coordinates are 
\begin{eqnarray}
	\label{eq:RNLag2fields}
    && \mathcal{K}_{\rm RNC} \; = \; \frac{1}{2} (\partial_{\mu} \tilde{\chi})^2 + \frac{1}{2} (\partial_{\mu} \tilde{\rho})^2 + \frac{1}{2} (\partial_{\mu} \tilde{\xi})^2 + \frac{1}{2} (\partial_{\mu} \tilde{\sigma})^2  - \frac{1}{6\alpha} \left(\tilde{\rho} \, \partial_{\mu} \tilde{\sigma} -  \tilde{\sigma} \, \partial_{\mu} \tilde{\rho} \right) (\tilde{\xi} \, \partial_{\mu} \tilde{\chi} -  \tilde{\chi} \, \partial_{\mu} \tilde{\xi} )~\nonumber \\    
   && - \frac{1}{18 \alpha} (\tilde{\rho} \, \partial_{\mu} \tilde{\rho} + \tilde{\sigma} \, \partial_{\mu} \tilde{\sigma} ) (\tilde{\xi} \, \partial^{\mu} \tilde{\xi} + \tilde{\chi} \, \partial^{\mu} \tilde{\chi}) + \frac{1}{9\alpha} \left( \left(\tilde{\rho} \, \partial_{\mu} \tilde{\sigma}  - \tilde{\sigma} \, \partial_{\mu} \tilde{\rho}  \right)^2 + (\tilde{\xi} \, \partial_{\mu} \tilde{\chi}  - \tilde{\chi} \, \partial_{\mu} \tilde{\xi} )^2 \right)  ~\nonumber \\
    && + \frac{1}{36\alpha} \left(((\partial_{\mu} \tilde{\xi})^2+ (\partial_{\mu} \tilde{\chi})^2) (\tilde{\rho}^2 + \tilde{\sigma}^2) +((\partial_{\mu} \tilde{\rho})^2 +
    (\partial_{\mu} \tilde{\sigma})^2)(\tilde{\chi}^2 + \tilde{\xi}^2)
    \right)    \, .
\end{eqnarray}

As discussed in the previous section, instead of using the vielbeins, the fields $(\chi, \rho, \xi, \sigma)$ can be transformed to Riemann normal coordinates $(\tilde{\chi}, \tilde{\rho}, \tilde{\xi}, \tilde{\sigma})$ via the relations~(\ref{rnctrans1}) that make the metric diagonal. Performing additional transformations that rescale the fields $\tilde \chi \rightarrow \tilde{G}_{\tilde{\chi} \tilde{\chi}}^{-1/2} \tilde{\chi}$, $\tilde \rho \rightarrow \tilde{G}_{\tilde{\rho} \tilde{\rho}}^{-1/2} \tilde{\rho}$, $\tilde \xi \rightarrow \tilde{G}_{\tilde{\xi}\tilde{\xi}}^{-1/2} \tilde{\xi}$, and $\tilde \sigma \rightarrow \tilde{G}_{\tilde{\sigma} \tilde{\sigma}}^{-1/2} \tilde{\sigma}$, we then obtain up to quartic order
\begin{eqnarray}
\label{rnctrans2fieldfulla}
  \chi & = & \frac{\sqrt{\frac{2}{3}}v}{\sqrt{\alpha}} \tilde{\chi} +\frac{v}{3\alpha}(\tilde{\chi}^2- \tilde{\xi}^2)+ \frac{v}{9\sqrt{6}\alpha^{3/2}} (2\tilde{\chi}^3-\tilde{\chi}(\tilde{\rho}^2+\tilde{\sigma}^2)-10\tilde{\xi}^2 \tilde{\chi})~\nonumber \\
   &+& \frac{v}{54\alpha^2}\left(\tilde{\chi}^4+5\tilde{\xi}^4-18\tilde{\xi}^2 \tilde{\chi}^2+2(\tilde{\xi}^2-\tilde{\chi}^2)(\tilde{\rho}^2+\tilde{\sigma}^2)\right)\,, \\
 \rho & = & \frac{\sqrt[4]{2}\sqrt{v}}{\sqrt{\alpha}} \tilde{\rho}+\frac{\sqrt{v}}{\sqrt[4]{2}\sqrt{3}\alpha }(\tilde{\rho } \tilde{\chi }-\tilde{\xi } \tilde{\sigma })-\frac{\sqrt[4]{2}\sqrt{v} }{18 \alpha ^{3/2}}(\tilde{\rho}^3-2\tilde{\rho} \tilde{\chi }^2+\tilde{\rho} \tilde{\sigma}^2+4\tilde{\rho}\tilde{\xi}^2+6\tilde{\xi} \tilde{\sigma} \tilde{\chi})~\nonumber\\
 &+&  \frac{\sqrt{v} }{18 \sqrt[4]{2} \sqrt{3} \alpha ^2}\left(\tilde{\rho} \tilde{\chi}^3-2  \tilde{\rho}^3 \tilde{\chi}-2\tilde{\rho} \tilde{\sigma}^2\tilde{\chi}+2 \tilde{\xi}\tilde{\sigma}^3+5\tilde{\xi}^3\tilde{\sigma}-7\tilde{\xi}\tilde{\sigma}\tilde{\chi}^2-11\tilde{\xi}^2\tilde{\rho}\tilde{\chi}+2\tilde{\xi}\tilde{\rho}^2\tilde{\sigma} \right),\qquad\\ 
\xi & = & \frac{\sqrt{\frac{2}{3}}v}{\sqrt{\alpha }}\tilde{\xi}+\frac{2 v }{3 \alpha}\tilde{\xi} \tilde{\chi}-\frac{v} {9 \sqrt{6} \alpha^{3/2}} (\tilde{\xi } (\tilde{\rho}^2+\tilde{\sigma}^2)+4\tilde{\xi}^3-8\tilde{\xi}\tilde{\chi}^2)~\nonumber\\  
&+&\frac{2v}{27 \alpha ^2}(2\tilde{\xi} \tilde{\chi}^3-4\tilde{\xi}^3 \tilde{\chi}-\tilde{\xi}  \tilde{\chi}(\tilde{\rho}^2+\tilde{\sigma}^2))
\, , \\ 
\sigma & = & \frac{\sqrt[4]{2} \sqrt{v}}{\sqrt{\alpha }}\tilde{\sigma} +\frac{\sqrt{v} }{\sqrt[4]{2} \sqrt{3} \alpha } (\tilde{\xi} \tilde{\rho}+\tilde{\sigma}\tilde{\chi}) -\frac{\sqrt[4]{2}\sqrt{v} }{18 \alpha ^{3/2}}(\tilde{\sigma}^3-2\tilde{\sigma} \tilde{\chi }^2+\tilde{\sigma}\tilde{\rho}^2 +4\tilde{\sigma}\tilde{\xi}^2-6\tilde{\xi} \tilde{\rho} \tilde{\chi}) ~\nonumber\\
&+&\frac{\sqrt{v} }{18 \sqrt[4]{2} \sqrt{3} \alpha ^2}\left(\tilde{\sigma} \tilde{\chi}^3-2  \tilde{\sigma}^3 \tilde{\chi}-2\tilde{\sigma} \tilde{\rho}^2\tilde{\chi}-2 \tilde{\xi}\tilde{\rho}^3-5\tilde{\xi}^3\tilde{\rho}+7\tilde{\xi}\tilde{\rho}\tilde{\chi}^2-11\tilde{\xi}^2\tilde{\sigma}\tilde{\chi}-2\tilde{\xi}\tilde{\rho}\tilde{\sigma}^2 \right).\nonumber \\ 
    \label{rnctrans2fieldfulld}
\end{eqnarray}
These field transformations make the metric diagonal and canonical. Using the Lagrangian (\ref{lag2fields}), we then recover the kinetic terms (\ref{eq:RNLag2fields}).

We also show how to transform the Lagrangian~(\ref{lag2fields}) to K\"ahler normal coordinates. Assuming the complex field decomposition, $T(x) = w+t(x)$ and $\phi = r+f(x)$,  where $t(x),f(x)$ are the field fluctuations and $w,r$ are the field VEVs, the K\"ahler metric~(\ref{kahmetric2fields}) becomes
\begin{equation}
	\label{kahmetric2fieldsvevs}
    K_{I \bar{J}}(w) \; = \; 
    \begin{pmatrix}
    \frac{3 \alpha}{(w + \bar{w})^2} & 0 \\
    0 & \frac{\alpha}{w + \bar{w}} 
    \end{pmatrix}
    \, ,
\end{equation}
where the field VEV $r = 0$.
The complex vielbein is given by
\begin{equation}
	\label{twofieldcomvielb}
    e(w, \bar{w}) \; = \; \bar{e} (w, \bar{w}) \; \; = \; 
    \begin{pmatrix}
    \frac{w + \bar{w}}{\sqrt{3 \alpha}} & 0 \\
    0 & \frac{\sqrt{w + \bar{w}}}{\sqrt{\alpha}}
    \end{pmatrix}
    \, ,
\end{equation}
which ensures that the K\"ahler metric is flat and satisfies the constraint $K_{I \bar{J}} e_{\alpha}{}^{I} \bar{e}_{\bar{\beta}}{}^{\bar{J}} = \delta_{\alpha \bar{\beta}}$.

Using expression~(\ref{kahcurvtens}), the non-zero components of the K\"ahler curvature tensor are
\begin{eqnarray}
    R_{t\bar{t}t\bar{t}} & = & \frac{6 \alpha}{(w + \bar{w})^4} \, , \qquad R_{f\bar{f}f\bar{f}} \; = \; \frac{2 \alpha}{3(w+\bar{w})^2}\, ,~\nonumber \\
    R_{t\bar{t}f\bar{f}} & = & R_{t\bar{f}f\bar{t}} \; = \; R_{f \bar{t} t \bar{f}} \; = \; R_{f \bar{f} t \bar{t}} \; = \;  \frac{\alpha}{(w + \bar{w})^3} \, .
\end{eqnarray}
In KNC, the Riemann tensor is obtained from  the expression $R_{\alpha \bar{\beta} \gamma \bar{\delta}} = e_{\alpha}{}^{I} \bar{e}_{\bar{\beta}}{}^{\bar{J}} e_{\gamma}{}^{\bar{K}} \bar{e}_{\bar{\delta}}{}^{\bar{L}} R_{I \bar{J} K \bar{L}}$, using the complex vielbein~(\ref{twofieldcomvielb}) to give
\begin{eqnarray}
    R_{\hat{t} \hat{\bar{t}} \hat{t} \hat{\bar{t}}} & = & R_{\hat{f} \hat{\bar{f}} \hat{f} \hat{\bar{f}}} \; = \; \frac{2}{3\alpha} \, , \\
    R_{\hat{t} \hat{\bar{t}} \hat{f} \hat{\bar{f}}} & = & R_{\hat{t} \hat{\bar{f}} \hat{f} \hat{\bar{t}}} \; = \; R_{\hat{f} \hat{\bar{t}} \hat{t} \hat{\bar{f}}} \; = \; R_{\hat{f} \hat{\bar{f}} \hat{t} \hat{\bar{t}}} \; = \; \frac{1}{3 \alpha} \, ,
\end{eqnarray}
where the hatted coordinates refer to the K\"ahler normal coordinate basis. Note that there is no longer any dependence on the field VEV.

Using the complex vielbeins together with a holomorphic transformation~(\ref{eq:transkah1}), and the field rescalings $\hat{t} \rightarrow \hat{G}_{\hat{t} \hat{\bar{t}}}^{-1/2}\hat{t}$, $\hat{f} \rightarrow \hat{G}_{\hat{f} \hat{\bar{f}}}^{-1/2}\hat{f}$, the field transformations, assuming ${\bar w} = w$, are given by
\begin{eqnarray}
	\label{holtrans2}
    t & =& \frac{2w  }{\sqrt{3 \alpha}}\hat{t} + \frac{2w  }{3 \alpha}\hat{t}^2 + \frac{2w }{3 \sqrt{3} \alpha^{3/2}} \hat{t}^3 + \frac{2 w }{9 \alpha^2}\hat{t}^4 + \ldots \; = \; 2 w \sum_{n=1}^{\infty} \left(\frac{\hat{t}}{\sqrt{3 \alpha}}  \right)^n \, , \\
    f & = & \frac{\sqrt{2 w} }{\sqrt{\alpha}}\hat{f} + \frac{\sqrt{2w} }{\sqrt{3} \alpha}\hat{f} \hat{t} + \frac{\sqrt{2w} }{3 \alpha^{3/2}}\hat{f} \hat{t}^2 + \frac{\sqrt{2w} }{3 \sqrt{3} \alpha^2} \hat{f} \hat{t}^3+ \ldots = \sqrt{2w} \hat{f} \sum_{n = 1}^{\infty} \frac{1}{\sqrt{\alpha}^n} \left(\frac{\hat{t}}{\sqrt{3}} \right)^{n-1}.\nonumber\\
    \label{holtrans3}
\end{eqnarray}
These field transformations make the K\"ahler metric diagonal and canonical. Substituting Eqs.~\eqref{holtrans2} and \eqref{holtrans3} into the Lagrangian~(\ref{lag2fields}), one obtains the kinetic terms in K\"ahler normal coordinates
\begin{equation}
    \mathcal{K}_{\rm{KNC}} = \partial_{\mu} \hat{t} \partial^{\mu} \hat{\bar{t}} + \partial_{\mu} \hat{f} \partial^{\mu} \hat{\bar{f}} + \frac{2}{3\alpha} \left(\hat{t} \hat{\bar{t}} \partial_{\mu} \hat{t} \partial^{\mu} \hat{\bar{t}} + \hat{f} \hat{\bar{f}} \partial_{\mu} \hat{f} \partial^{\mu} \hat{\bar{f}} \right)
     +\frac{1}{3\alpha} \left(\hat{t} \partial_{\mu} \hat{f}  + \hat{f} \partial_{\mu} \hat{t} \right) \left(\hat{\bar{t}} \partial_{\mu} \hat{\bar{f}}  + \hat{\bar{f}} \partial_{\mu} \hat{\bar{t}} \right)\, ,
     \label{eq:knc2comp}
\end{equation}
or in terms of the real fields, 
\begin{eqnarray}
	\label{eq:KNLag2fields2}
    \mathcal{K}_{\rm{KNC}} &= & \frac{1}{2} (\partial_{\mu} \hat{\chi})^2 + \frac{1}{2} (\partial_{\mu} \hat{\rho})^2 + \frac{1}{2} (\partial_{\mu} \hat{\xi})^2 + \frac{1}{2} (\partial_{\mu} \hat{\sigma})^2 ~\nonumber \\
  &  + &\frac{1}{6\alpha} \left( ((\partial_{\mu} \hat{\chi})^2 + (\partial_{\mu} \hat{\xi})^2) (\hat{\chi}^2 + \hat{\xi}^2) + ((\partial_{\mu} \hat{\rho})^2 + (\partial_{\mu} \hat{\sigma})^2) (\hat{\rho}^2 + \hat{\sigma}^2)
\right.~\nonumber \\
&&\left.  
\qquad+ (\partial_{\mu} \hat{\rho} \partial^{\mu} \hat{\chi} +\partial_{\mu} \hat{\sigma} \partial^{\mu} \hat{\xi})
(\hat{\rho}\hat{\chi}+\hat{\sigma}\hat{\xi})+(\partial_{\mu} \hat{\rho} \partial^{\mu} \hat{\xi}-\partial_{\mu} \hat{\sigma} \partial^{\mu} \hat{\chi})
(\hat{\rho}\hat{\xi}-\hat{\sigma}\hat{\chi})
 \right)~\nonumber \\
  & +& \frac{1}{12\alpha} \left( ((\partial_{\mu} \hat{\chi})^2 + (\partial_{\mu} \hat{\xi})^2) (\hat{\rho}^2 + \hat{\sigma}^2) + ((\partial_{\mu} \hat{\rho})^2 + (\partial_{\mu} \hat{\sigma})^2) (\hat{\chi}^2 + \hat{\xi}^2)  \right)\, ,
\end{eqnarray}
where the complex fields are decomposed as $\hat{t} = \frac{1}{\sqrt{2}}\left(\hat{\chi} + i \hat{\xi} \right)$ and $\hat{f} = \frac{1}{\sqrt{2}}\left(\hat{\rho} + i \hat{\sigma} \right)$.
From Eq.~(\ref{eq:knc2comp}), we see that there are four combinations of the complex fields with non-vanishing amplitudes, given by
\begin{align}
    &A^{t \, t \, \bar{t}\, \bar{t}}_{4, kin} \; = \; A^{f f \bar{f} \bar{f}}_{4, kin} \; = \; \frac{2}{3\alpha} s_{12} \, ,~\nonumber \\
    &A^{f \, t \, \bar{f}\, \bar{t}}_{4, kin} \; = \; A^{t \, f \, \bar{t}\, \bar{f}}_{4, kin} \; = \; \frac{1}{3\alpha} s_{12} \,.
\end{align}

Similarly, from Eqs.~(\ref{eq:RNLag2fields}) and (\ref{eq:KNLag2fields2}), we can write the 
four-point scattering amplitudes arising from the kinetic terms in terms of real fields:
\begin{align}
    &A^{\chi \chi \xi \xi}_{4, kin} \; = \; A^{\xi \xi \chi \chi}_{4, kin} \; = \; A^{\rho \rho \sigma \sigma}_{4, kin} \; = \; A^{\sigma \sigma \rho \rho}_{4, kin} \; = \; -\frac{2}{3\alpha}s_{12} \, , ~\nonumber\\
    &A^{\chi \xi \chi \xi}_{4, kin} \; = \; A^{\xi \chi \xi \chi}_{4, kin} \; = \; A^{\rho \sigma \rho \sigma}_{4, kin} \; = \; A^{\sigma \rho \sigma \rho}_{4, kin} \; = \; -\frac{2}{3\alpha}s_{13}\, ,~\nonumber \\
    &A^{\chi \xi \xi \chi}_{4, kin} \; = \; A^{\xi \chi \chi \xi}_{4, kin} \; = \; A^{\rho \sigma \sigma \rho}_{4, kin} \; = \; A^{\sigma \rho \rho \sigma}_{4, kin} \; = \; -\frac{2}{3\alpha}s_{14} \, ,~\nonumber \\
    &A^{\chi \chi \rho \rho}_{4, kin} \; = \; A^{\rho \rho \chi \chi}_{4, kin} \; = \; A^{\chi \chi \sigma \sigma}_{4, kin} \; = \; A^{\sigma \sigma \chi \chi}_{4, kin} \; = \; -\frac{1}{6\alpha}s_{12}\,, ~\nonumber\\
    &A^{\chi  \rho  \chi \rho}_{4, kin} \; = \; A^{\rho \chi \rho \chi}_{4, kin} \; = \; A^{\chi  \sigma \chi \sigma}_{4, kin} \; = \; A^{\sigma  \chi \sigma \chi}_{4, kin} \; = \; -\frac{1}{6\alpha}s_{13}\,, ~\nonumber\\
    &A^{\chi  \rho \rho \chi}_{4, kin} \; = \; A^{\rho  \chi \chi \rho}_{4, kin} \; = \; A^{\chi \sigma \sigma \chi}_{4, kin} \; = \; A^{\sigma \chi \chi \sigma}_{4, kin} \; = \; -\frac{1}{6\alpha}s_{14}\,.
\end{align}
As before, we omit the hats or tildes in these expressions because the scattering amplitudes coincide in Riemann and K\"ahler normal coordinates. 

\subsubsection{General Two-Derivative Interactions}
\begin{figure}[ht!]
    \centering
    \includegraphics[width=0.76\textwidth]{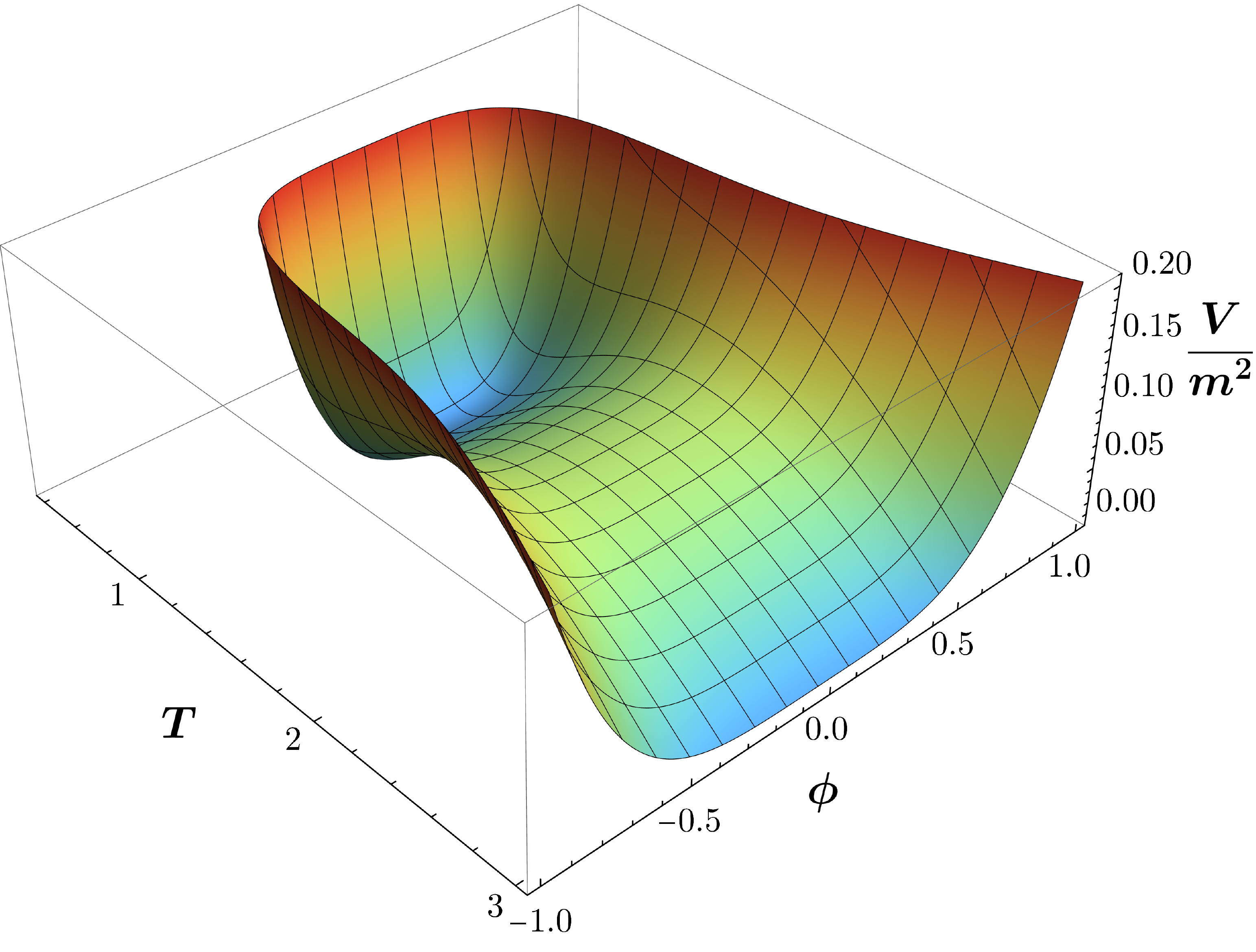}
    \caption{The scalar potential~(\ref{eq:3deffpot}) as a function of $T$ and $\phi$, where $\bar{T} = T$, $\bar{\phi} = \phi$, and $\lambda = m$. The global minimum is located at $T = \bar{T} = 1$ and $\phi = \bar{\phi} = 0$, which in terms of real scalar fields corresponds to $(X, Y, P, S) = (\sqrt{2}, 0, 0, 0)$.}
    \label{fig:3dplot1}
\end{figure}

Finally, in this section, we consider a two-field superpotential obtained from adding a cubic term $\phi^3$ to the quadratic superpotential (\ref{eq:spT})
\begin{equation}
    W(T,\phi) \; = \; m (T - 1)^2 + \lambda \phi^3 \, .
\end{equation}
Using Eq.~(\ref{eq:effpotsugra}) with the no-scale K\"ahler potential~(\ref{kah1}), the scalar potential is then given by
\begin{equation}
    \label{eq:3deffpot}
    V \; = \; \frac{1}{(T +\bar{T} - \frac{|\phi|^2}{3})} \left(\frac{2}{3}m^2 |T - 1|^2 (6 - T - \bar{T}) + 9 \lambda^2 |\phi|^4 \right) \, ,
\end{equation}
where $\alpha = 1$. Expanding the complex fields as $T = \frac{X + i Y}{\sqrt{2}}$ and $\phi = \frac{P + i S}{\sqrt{2}}$, we obtain the scalar potential in terms of the real fields
\begin{equation}
    V \; = \; \frac{81 \lambda ^2 \left(P^2+S^2\right)^2-12 m^2 \left(X \left(X \left(\sqrt{2} X-10\right)+\sqrt{2} \left(Y^2+14\right)\right)-6 \left(Y^2+2\right)\right)}{\left(P^2+S^2-6 \sqrt{2} X\right)^2} \, .
\end{equation}
One can easily find that the global minimum of the potential is located at $(X, Y, P, S) = (\sqrt{2}, 0, 0, 0)$. We plot the scalar potential~(\ref{eq:3deffpot}) along the real directions $\bar{T} = T$ and $\bar{\phi} = \phi$ with $m = \lambda$. We illustrate it in Fig.~\ref{fig:3dplot1} and show that the global minimum is located at $T = \bar{T} = 1$ and $\phi = \bar{\phi} = 0$, which corresponds to $(X, Y, P, S) = (\sqrt{2}, 0, 0, 0)$. Along the $\phi = 0$ direction, the one-dimensional potential is the same as that in Fig.~\ref{fig:V2} which exhibits a minimum and maximum (here the latter is a saddle point). 

If we decompose the fields $X(x) = v + \chi(x)$, $Y(x) = u + \xi(x)$, $P(x) = p+\rho(x)$, and $S(x) = s+\sigma(x)$, and use the field transformations to Riemann normal coordinates~(\ref{rnctrans2fieldfulla})-(\ref{rnctrans2fieldfulld}), the potential to quartic order is given by
\begin{eqnarray}
\label{2fieldVRNC}
 &&V_{\rm{RNC}} = \frac{4}{9} m^2 (\tilde{\chi }^2+\tilde{\xi }^2)+\frac{9}{4} \lambda ^2(\tilde{\rho }^2 +\tilde{\sigma }^2)^2-\frac{2}{3} \sqrt{\frac{2}{3}} m^2 (\tilde{\xi }^2 \tilde{\chi }+\tilde{\chi }^3)~\nonumber\\
 &&\qquad\qquad+\frac{20}{81} m^2 (\tilde{\chi }^2+\tilde{\xi }^2)^2+\frac{8}{81} m^2 (\tilde{\chi }^2+\tilde{\xi }^2)(\tilde{\rho }^2+\tilde{\sigma }^2)\,,
\end{eqnarray}
where the VEV values $(v, u, p, s) = (\sqrt{2}, 0, 0, 0)$. Similarly, using the field transformations to K\"ahler normal coordinates~(\ref{holtrans2})(\ref{holtrans3}), we obtain
\begin{eqnarray}
\label{2fieldVKNC}
 &&V_{\rm{KNC}} = \frac{4}{9} m^2 (\tilde{\chi }^2+\tilde{\xi }^2)+\frac{9}{4} \lambda ^2(\tilde{\rho }^2 +\tilde{\sigma }^2)^2-\frac{2}{3} \sqrt{\frac{2}{3}} m^2 (\tilde{\xi }^2 \tilde{\chi }+\tilde{\chi }^3)~\nonumber\\
 &&\qquad\qquad+\frac{8}{27} m^2 (\tilde{\chi }^2+\tilde{\xi }^2)^2+\frac{4}{27} m^2 (\tilde{\chi }^2+\tilde{\xi }^2)(\tilde{\rho }^2+\tilde{\sigma }^2)\,.
\end{eqnarray}
The two potentials \eqref{2fieldVRNC} and \eqref{2fieldVKNC} only differ in the quartic interactions.
In both cases, the mass matrix is diagonal and the masses are given by
\begin{equation}
    m_{\chi}^2 \; = \; \frac{8m^2}{9}, \qquad m_{\rho}^2 \; = \; 0 \qquad m_{\xi}^2 \; = \; \frac{8m^2}{9}, \qquad m_{\sigma}^2 \; = \; 0 \, ,
\end{equation}
where we again note that the scalar masses coincide in RNC and KNC.

As before, we evaluate selected four-point amplitudes in RNC and KNC in the nonrelativistic limit choosing the Mandelstam variables in a similar fashion as described in the earlier examples.  The amplitudes for a selection of scattering channels are summarized in Table~\ref{tab:Pot4pt3} where the amplitudes are evaluated at the minimum, $T = {\bar T} = 1$ and $\phi = {\bar \phi} = 0$. Once again, we see that for each channel, the total amplitudes are invariant with respect to 
the choice of coordinates even though the contributions from kinetic and potential terms differ.

\begin{table}[t!]
    \centering
    \begin{tabular}{c|c|c||c||c|c||c||}
    \cline{2-7}
    & \multicolumn{3}{|c||}{\bf{RNC}}&
    \multicolumn{3}{|c||}{\bf{KNC}}\\
    \cline{2-7}
    \hhline{~======}
      & $\mathcal{K}$ & $V$ & Total &  $\mathcal{K}$ & $V$ & Total\\
     \hline
      \multicolumn{1}{|c|}{$\chi \chi \rightarrow \chi \chi$} & $0$ & $\frac{380m^2}{27}$ & $\frac{380m^2}{27}$ &  $\frac{32m^2}{27}$ & $\frac{116m^2}{9}$ & $\frac{380m^2}{27}$ \\
         \hline
         \multicolumn{1}{|c|}{$\rho \rho \rightarrow \rho \rho$} & $0$ & $-54\lambda^2$ & $-54\lambda^2$ & $0$ & $-54\lambda^2$ & $-54\lambda^2$  \\
         \hline
       \multicolumn{1}{|c|}  {$\xi \xi \rightarrow \xi \xi$} & $0$ & $-\frac{100m^2}{27}$ & $-\frac{100m^2}{27}$ & $\frac{32m^2}{27}$ & $-\frac{44m^2}{9}$ & $-\frac{100m^2}{27}$ \\
      \hline
      \multicolumn{1}{|c|}{$\chi \chi \rightarrow \rho \rho$} &$-\frac{40m^2}{81}$ &$-\frac{32m^2}{81}$ & $-\frac{8m^2}{9}$ &$-\frac{8m^2}{27}$ &$-\frac{16m^2}{27}$ & $-\frac{8m^2}{9}$\\
         \hline
         \multicolumn{1}{|c|}{$\chi \chi \rightarrow \xi \xi$} & $-\frac{128m^2}{81}$ & $-\frac{52m^2}{81}$ & $-\frac{20m^2}{9}$ & $-\frac{32m^2}{27}$ & $-\frac{28m^2}{27}$ & $-\frac{20m^2}{9}$ \\
           \hline 
         \multicolumn{1}{|c|}{$\chi \rho \rightarrow \chi \rho$} & $\frac{8m^2}{81}$ & $-\frac{32m^2}{81}$ & $-\frac{8m^2}{27}$ & $\frac{8m^2}{27}$ & $-\frac{16m^2}{27}$  & $-\frac{8m^2}{27}$ \\
           \hline 
           \multicolumn{1}{|c|}{$\chi \xi \rightarrow \chi \xi$} & $\frac{64m^2}{81}$ &$\frac{236m^2}{81}$ & $\frac{100m^2}{27}$ & $\frac{32m^2}{27}$ & $\frac{68m^2}{27}$ & $\frac{100m^2}{27}$ \\\hline
  \multicolumn{1}{|c|}{$\rho \rho \rightarrow \sigma \sigma$} & 0 & $-18\lambda^2$ & $-18\lambda^2$ &  0 & $-18\lambda^2$ & $-18\lambda^2$ \\
           \hline 
\multicolumn{1}{|c|}{$\chi \rho \rightarrow \xi \sigma$} & $-\frac{16m^2}{27}$ & $0$ & $-\frac{16m^2}{27}$ &  $-\frac{16m^2}{27}$ & 0 & $-\frac{16m^2}{27}$ \\
           \hline 
\multicolumn{1}{|c|}{$\chi \sigma \rightarrow \xi \rho$} & $\frac{16m^2}{27}$ & $0$ & $\frac{16m^2}{27}$ &  $\frac{16m^2}{27}$ & 0 & $\frac{16m^2}{27}$ \\
           \hline 
    \end{tabular} 
    \caption{Four-point scattering amplitudes in the nonrelativistic limit (where we assume the energy, $E_{\rho,\sigma}=m_{\chi,\xi}$, if one of the incoming particles is massless). As expected, the amplitudes match between the RNC and KNC when evaluated at the minimum (i.e. $(T,\phi)=(1,0$)). Note that only distinct scattering amplitudes are listed in the table, which are related to other amplitudes not shown e.g., $A_4(\sigma \sigma \rightarrow \sigma \sigma)=A_4(\rho \rho \rightarrow \rho \rho)$.}
    \label{tab:Pot4pt3}
\end{table}

\begin{figure}[ht!]
    \centering
    \includegraphics[width=0.8\columnwidth]{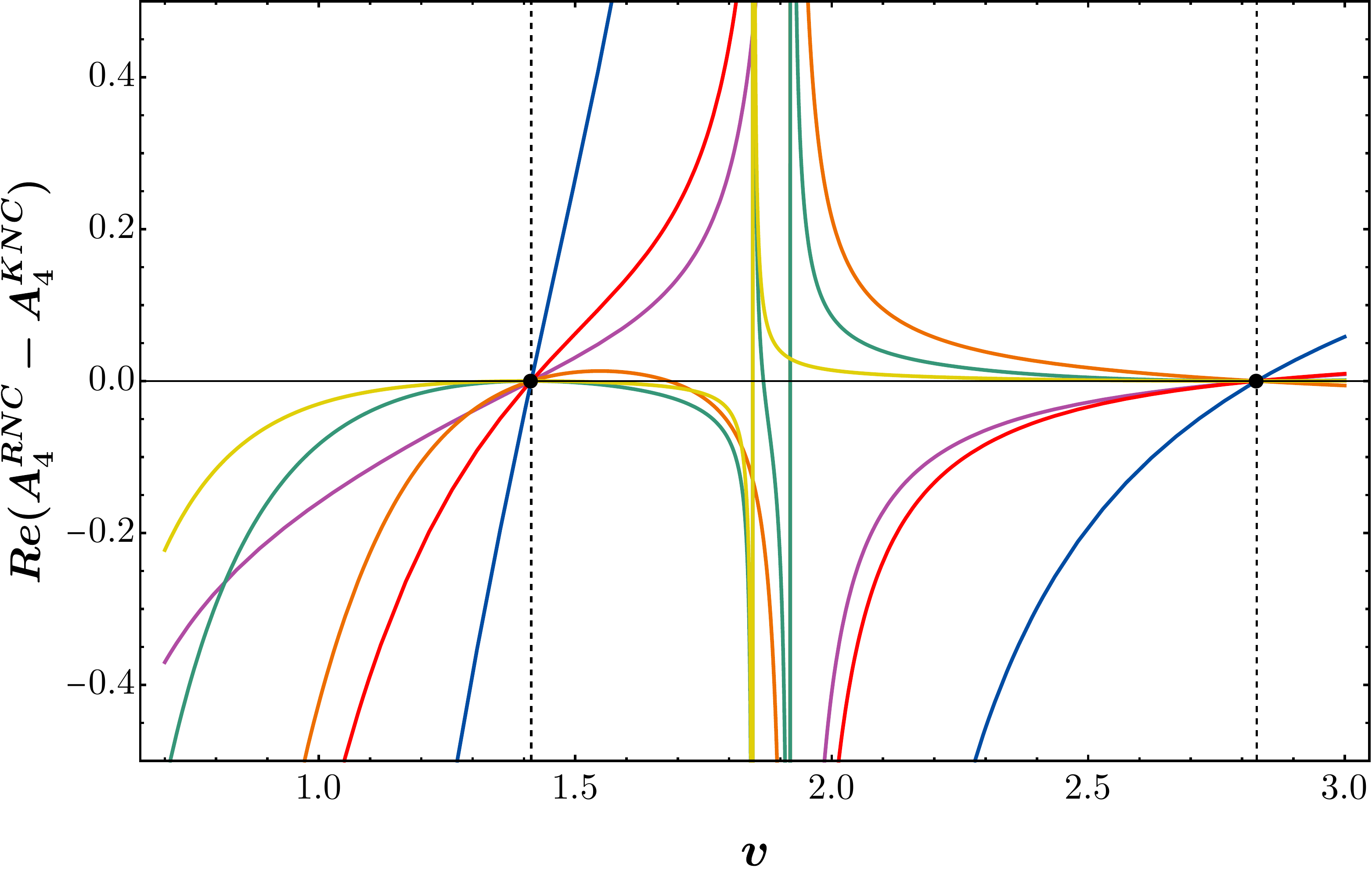}
    \caption{The difference between the real values of RNC and KNC four-point amplitudes for the potential \eqref{eq:3deffpot} as a function of $v$ for different channels (blue: $\chi \chi \rightarrow \chi \chi$, purple: $\xi \xi \rightarrow \xi \xi$, orange: $\chi \chi \rightarrow \xi \xi$, red: $\xi \xi \rightarrow \chi \chi$, green: $\rho \rho \rightarrow \rho \rho$, yellow: $\rho \rho \rightarrow \sigma \sigma$). We assume that $\alpha = 1$ and $m = \lambda = 1$. The amplitudes only agree at $v = \sqrt{2}$ (global minimum) and $v= 2\sqrt{2}$ (saddle point).
    }
    \label{fig:twocomfieldampdiffs}
\end{figure}

The difference in the scattering amplitudes between RNC and KNC for the two complex field case is shown for six different scattering channels as a function of the VEV of $T$ in Fig.~\ref{fig:twocomfieldampdiffs}. We see clearly that the difference vanishes at both extrema, $v=\sqrt{2}$ (minimum) and $v=2 \sqrt{2}$ (saddle point). 

\section{Discussion and Conclusions}
\label{summary}

In this work, we have presented detailed calculations for scattering amplitudes in ${\cal N} = 1$ supergravity theories.
 Beyond minimal supergravity with $K_{i{\bar j}} = \delta_{i{\bar j}}$, due to the complexity of non-minimal kinetic terms, we have argued that the calculation of scattering amplitudes is best done after field redefinitions to either Riemann normal coordinates (with $2N$ real fields associated with $N$ chiral multiplets) or K\"ahler normal coordinates (with $N$ complex fields). We have provided explicit expressions for the contributions to the amplitudes arising from kinetic and potential terms that depend on geometric quantities. This is consistent with the known invariance property of the $S$-matrix under field redefinitions~\cite{Chisholm:1961tha,Kamefuchi:1961sb,Arzt:1993gz}.

 We have also demonstrated that 
 the amplitudes are invariant under the field redefinitions only at the extrema of the potential. Away from these points, invariance may be maintained only if all contributions from tadpole (linear) terms are included in the calculation. In general, we know of no method to include these contributions which can proliferate into an infinite series of diagrams, except for simple examples, see e.g.  \cite{Dudas:2004nd}. We return to this question in future work \cite{dgov2}. It should also be noted that we have neglected the possibly dominant vacuum energy when calculating the scattering amplitudes at the maxima.  A more complete calculation of the scattering amplitudes should be done in a de Sitter background, but we expect that our qualitative results will remain unchanged.

 As specific examples, we have concentrated on no-scale supergravity models \cite{no-scale} with K\"ahler potentials given by Eqs.~(\ref{kah1}) and (\ref{v0}). We considered models with a single chiral superfield, $T$, as well as models with two chiral superfields that include matter, both with and without potential contributions to the Lagrangian. The calculation of the four-point scattering amplitudes is made much simpler by transforming to either RNC or KNC where cubic derivative terms are eliminated. Expressions for amplitudes beyond four-point can also be derived using our methods, by generalizing the results of Ref.~\cite{Cheung:2021yog}.

 While the calculations in this paper were performed at tree level, it would be interesting to extend the calculation to the loop level where we expect the scattering amplitudes to again depend on geometric quantities~\cite{Helset:2022tlf} and remain invariant only at potential extrema. Furthermore, generalizing the calculation to include fermions would lead to manifestly supersymmetric results in terms of superfields and remains another direction to explore.

 As we said at the outset, one of the motivations for this work is the SWGC. Testing specific models with respect to 
 this conjecture requires us to fully calculate the amplitudes in the no-scale context, as this is the framework expected to arise from string theory \cite{Witten}. As we have seen, testing the SWGC in this context is highly non-trivial, as the scalars generally have considerable mixing through their kinetic  terms as determined from the K\"ahler potential. In subsequent work,
 we will compare the amplitudes computed here with the gravitational amplitudes. We have already seen that for the example of a single complex field with no superpotential, the amplitudes for $\chi \chi \to \chi \chi$ and $\xi \xi \to \xi \xi$ vanish, violating the conjecture. We will also test models of inflation constructed from no-scale supergravity \cite{building}.
 However, we caution that the results found in this work indicate that the strong version of the SWGC may not be well founded as
 the calculation of amplitudes (without the inclusion of tadpoles) is not invariant under field redefinitions. 
 
\section*{Acknowledgements}
We would like to thank C. Cheung, M. A. G. Garc{\'i}a, A. Helset, D. Sutherland, and A. Vainshtein for useful discussions. The work of T.G and K.A.O.~was supported in part by DOE grant DE-SC0011842 at the University of Minnesota. The work of T.G. was performed in part at the Aspen Center for Physics, which is supported by National Science Foundation grant PHY-1607611. The work of S.V. was supported in part by DOE grant DE-SC0022148.

\end{document}